\documentclass[12pt,dvipsnames]{article}

\usepackage{scicite}

\usepackage{times}
\usepackage{graphicx}
\usepackage[T1]{fontenc}
\usepackage{siunitx}
\usepackage{chemformula}    
\usepackage{bm}        
\usepackage{amsfonts}  
\usepackage{amsmath}   
\usepackage{amssymb}   

\usepackage[normalem]{ulem}
\usepackage{xcolor}

\DeclareSIUnit\rydberg{Ry}

\newenvironment{sciabstract}{%
\begin{quote} \bf}
{\end{quote}}

\title{Quantum phase diagram of high-pressure hydrogen} 

\author
{Lorenzo Monacelli,$^{1\ast}$ Michele Casula,$^{2\ast}$ Kosuke Nakano,$^{3,4}$\\
Sandro Sorella,$^{4}$ Francesco Mauri$^{1\ast}$\\
\\
\normalsize{$^{1}$University of Rome, ``Sapienza'', Dipartimento di
  Fisica,}\\
\normalsize{Piaz.le Aldo Moro 5, 00185, Rome, Italy}\\ 
\normalsize{$^{2}$Institut de Min\'eralogie, de
  Physique des Mat\'eriaux et de Cosmochimie, Sorbonne Universit\'e,}\\
\normalsize{CNRS UMR7590, MNHN, 4 Place Jussieu, 75005, Paris, France}\\
\normalsize{$^{3}$Japan Advanced Institute of Science and Technology (JAIST),}\\
\normalsize{Asahidai 1-1, Nomi, Ishikawa 923-1292, Japan}\\
\normalsize{$^{4}$International School for Advanced Studies (SISSA), Via Bonomea 265, 34136, Trieste, Italy}\\
\normalsize{$^\ast$To whom correspondence should be addressed:}\\ \normalsize{ lorenzo92monacelli@gmail.com; francesco.mauri@uniroma1.it; michele.casula@impmc.upmc.fr}
}

\date{}

\begin{document} 

\baselineskip24pt

\maketitle 

\begin{sciabstract}
The interplay between electron correlation and nuclear quantum effects makes
our understanding of elemental hydrogen a formidable challenge.  
Here, we present the phase diagram of hydrogen and deuterium at low
temperatures and high-pressure ($P > \SI{300}{\giga\pascal})$ by
accounting for highly accurate electronic and nuclear enthalpies. We
evaluated internal electronic energies by diffusion quantum Monte
Carlo, while nuclear quantum motion and anharmonicity have been
included by the stochastic self-consistent harmonic
approximation. 
Our results show that the long-sought atomic metallic hydrogen, 
predicted to host room-temperature superconductivity, 
forms at \SI{577\pm10}{\giga\pascal} (\SI{640\pm 14}{\giga\pascal}
in deuterium). Indeed, anharmonicity pushes the stability of 
this phase towards pressures much larger
than previous theoretical estimates or attained experimental values.
Before atomization, molecular hydrogen transforms
from a conductive phase III to another metallic structure that is still molecular (phase VI)
at \SI{422\pm40}{\giga\pascal} (\SI{442\pm30}{\giga\pascal} in deuterium). 
We predict clear-cut signatures in optical spectroscopy and DC conductivity that can be used experimentally to distinguish between the two structural transitions. 
According to our findings, the experimental evidence of metallic hydrogen has so far been limited to
molecular phases.
\end{sciabstract}

In 1968, Ashcroft predicted that 
atomic 
metallic
hydrogen is a room
temperature superconductor\cite{AshcroftHydrogen}. During the last
fifty years, a lot of effort 
was
devised to synthesize atomic hydrogen in laboratory under stable
conditions. Nonetheless, the challenge proved more difficult than
expected.  
Solid hydrogen at high pressures exhibits a very rich phase diagram
with the presence of five different molecular phases,
labeled from I to
V\cite{mao1994ultrahigh,Howie_2012,DalladaySimpson2016}.
Recently, a new phase transition 
has been observed above \SI{420}{\giga\pascal} into a metallic 
state 
by
infrared (IR) absorption measurements\cite{Loubeyre2019observation}, i.e. phase VI.
However, Eremets \emph{et al.}\cite{Eremets2019_hydro} measured the Raman and reflectivity 
spectra of hydrogen up to \SI{480}{\giga\pascal} without incurring in any evidence of 
a sudden change in the sample. At even larger pressures,
Diaz and Silvera\cite{Dias2017} claimed to have synthesized the atomic metallic hydrogen from
reflectivity measurements at \SI{495}{\giga\pascal}, with reflectivity
data in good agreement with theoretical
predictions\cite{Borinaga_2018}, although the reliability of their
observation has been
questioned\cite{Goncharov2017,Silvera2017}.
Further uncertainties come from technical difficulties in determining
pressure at these extreme conditions, which could lead to a mismatch
of up to \SI{80}{\giga\pascal}\cite{Loubeyre2017}, jeopardizing the
possibility to reproduce results by independent studies. 

The structural characterization of these phases is challenging since
both neutron and X-ray diffraction experiments require sample sizes
non-compatible with pressures larger
than \SI{250}{\giga\pascal}\cite{Ji2019} in hydrostatic conditions. 
Numerical \emph{ab initio} simulations play consequently a crucial
role in understanding the phase diagram and can in principle address the following questions:
Have atomic metallic hydrogen been synthesized yet? At which pressure do
we expect to stabilize it?
What is the effect of isotope substitution? 

Nevertheless, also \emph{ab initio} approaches have been plagued so far by
severe limitations. Indeed, as hydrogen is
the lightest element, its nucleus is subject to huge quantum
fluctuations that can largely affect its structural properties. Indeed,
nuclear quantum effects have been shown to completely 
reshape the Born-Oppenheimer (BO) energy landscape in  
hydrogen-rich materials at
high-pressure\cite{Straus1977,Errea2015,ErreaNature2020},
invalidating the phase diagram obtained with classical
simulations.
Furthermore, many competing structures differ in enthalpy by less than
$\SI{1}{\milli\electronvolt}$ per atom in a broad range of
pressures\cite{Drummond2015}. This makes the identification of the
ground state very sensitive to  
approximations,
like the choice of 
the
exchange-correlation functional in density functional theory (DFT)
calculations. To overcome this issue, more sophisticated and accurate
theories 
are required, such as the quantum Monte Carlo (QMC) methods\cite{foulkes2001}.
For these reasons, the establishment of 
theoretical calculations fully accounting for both 
electron correlation energy and lattice anharmonicity at the same level of accuracy
is fundamental to determine the hydrogen phase diagram at such high
pressure.

To answer the aforementioned questions, we performed
hydrogen phase-diagram calculations at a methodological cutting edge,
by combining the highly accurate
description of electron correlation, within diffusion quantum Monte Carlo (DMC), 
and the anharmonic lattice optimization accounting for nuclear quantum
effects, within the stochastic self-consistent harmonic
approximation\cite{Errea2014,Bianco2017,Monacelli2018A,SCHA2021}
(SSCHA).  
DMC is a well-established framework that provides the most accurate
internal energies of solid
hydrogen\cite{Drummond2015,McMinis_2015,Azadi2017shissor,Monserrat_2018}. Here,
it has been coupled to SSCHA in a seamless fashion with the aim of
including both electronic and nuclear contributions in a
non-perturbative way. 
Indeed, SSCHA
can outperform other approximations to compute anharmonic phase
diagrams in hydrogen\cite{Azadi_2014,Drummond2015,Monserrat_2018}
thanks to its variational nature in determining free energy
differences and 
its
ability to relax 
atomic positions and lattice vectors.  
In our approach, SSCHA provides both vibrational energies and average
nuclear positions (centroids), calculated from nuclear quantum
fluctuations developing on the top of a DFT-BLYP\cite{BLYP} energy landscape. 
The resulting centroids crystal structure is used to compute 
DMC
electronic internal energies, that are combined with the 
SSCHA
zero-point
energies
of the anharmonic lattice.

\begin{figure*}
    \centering
    \includegraphics[width=\textwidth]{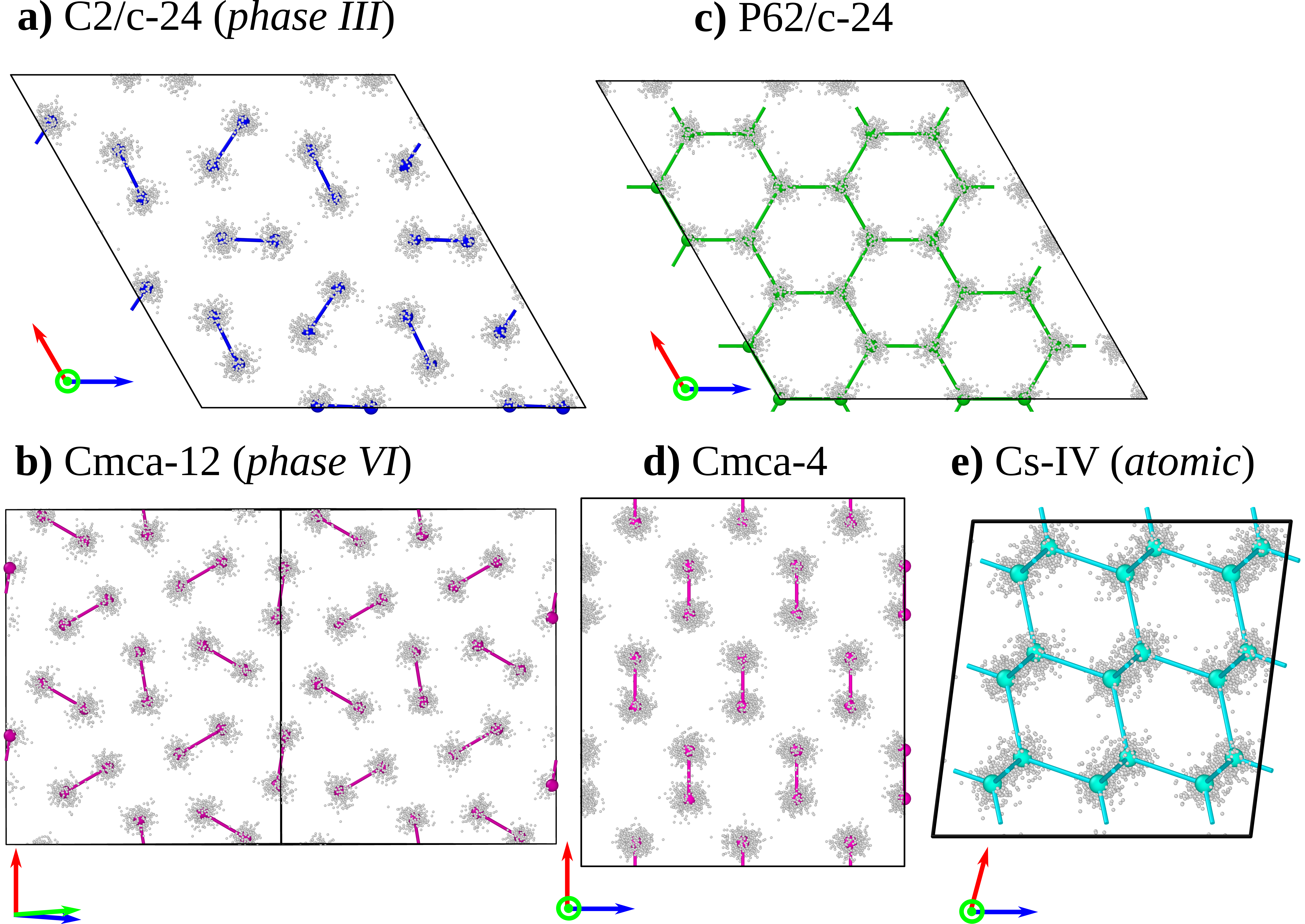}
    \caption{
    Structures considered for
      the low-temperature high-pressure phase diagram of hydrogen.
      Colored balls are the average centroid positions, sticks represent the \ch{H2} molecules,
      the cloud of smaller gray balls is a set of 250 configurations that sample the quantum probability distribution at \SI{0}{\kelvin}. All structures, apart from the atomic one, are made of layers, of which we report only one. P62/c-24 is made of alternating layers, one with atoms arranged in a honeycomb lattice (panel \textbf{b}), the other with molecular \ch{H2} 
      in a C2/c-24 arrangement 
      (not reported here).
  } 
    \label{fig:all_structures}
\end{figure*}

We modeled phase III as the monoclinic C2/c-24
structure\cite{Pickard_2007}\footnote{In line with previous
literature, we name phases with the symmetry group followed by the
number of atoms in the primitive cell.},
which is the best representative of this phase (Fig.~\ref{fig:all_structures}\textbf{a}). Indeed, 
not only is it
the most stable in its pressure range, but also it reproduces 
optical transmission and reflectivity, Raman and IR experimental data\cite{MonacelliNatPhys2020}.

Besides C2/c-24, we took into account
the Cmca-12 crystalline symmetry (Fig.~\ref{fig:all_structures}\textbf{b}), a new hexagonal structure with P62/c-24 symmetry (Fig.~\ref{fig:all_structures}\textbf{c}), and the Cmca-4 structure (Fig.~\ref{fig:all_structures}\textbf{d}) as the
most promising molecular 
geometries
for phase VI. 
Cmca-12 has firstly been suggested
by Pickard and Needs as an alternative candidate for phase
III\cite{Pickard_2007}, and more recently proposed as phase
VI\cite{McMinis_2015}.  
Cmca-4\cite{Pickard_2007} is the ground state in the harmonic DFT
phase diagram with common functionals (PBE and BLYP)  
over a large pressure range, but considerably disfavored by 
the most accurate QMC
internal energies\cite{Drummond2015,McMinis_2015}.
We discovered the new P62/c-24
structure by relaxing the symmetry constraint on the C2/c-24 with quantum anharmonic fluctuations above
\SI{320}{\giga\pascal} at the DFT-BLYP level of theory. It is
made of 
graphene-like sheets 
alternating
with molecular layers, conferring to the phase similar optical properties to graphite. It is a saddle-point of the BO energy landscape, stabilized by quantum fluctuations (see Supplementary Materials (SM)). However, it turns out that also this crystalline symmetry is disfavored by the QMC energies. 

Finally,
we 
simulated the atomic phase I4/amd-2, also named Cs-IV\cite{McMinis_2015} (Fig.~\ref{fig:all_structures}\textbf{e}).
According to DFT, it is the most stable atomic symmetry beyond the
molecular phases, and it is the one where room-temperature
superconductivity has been predicted\cite{Borinaga2017}.
In \figurename~\ref{fig:all_structures}, we report the centroid
positions of these structures at \SI{650}{\giga\pascal} with 
a visualization of the amplitude of quantum fluctuations. 

\begin{figure*}
    \centering
    \includegraphics[width=\textwidth]{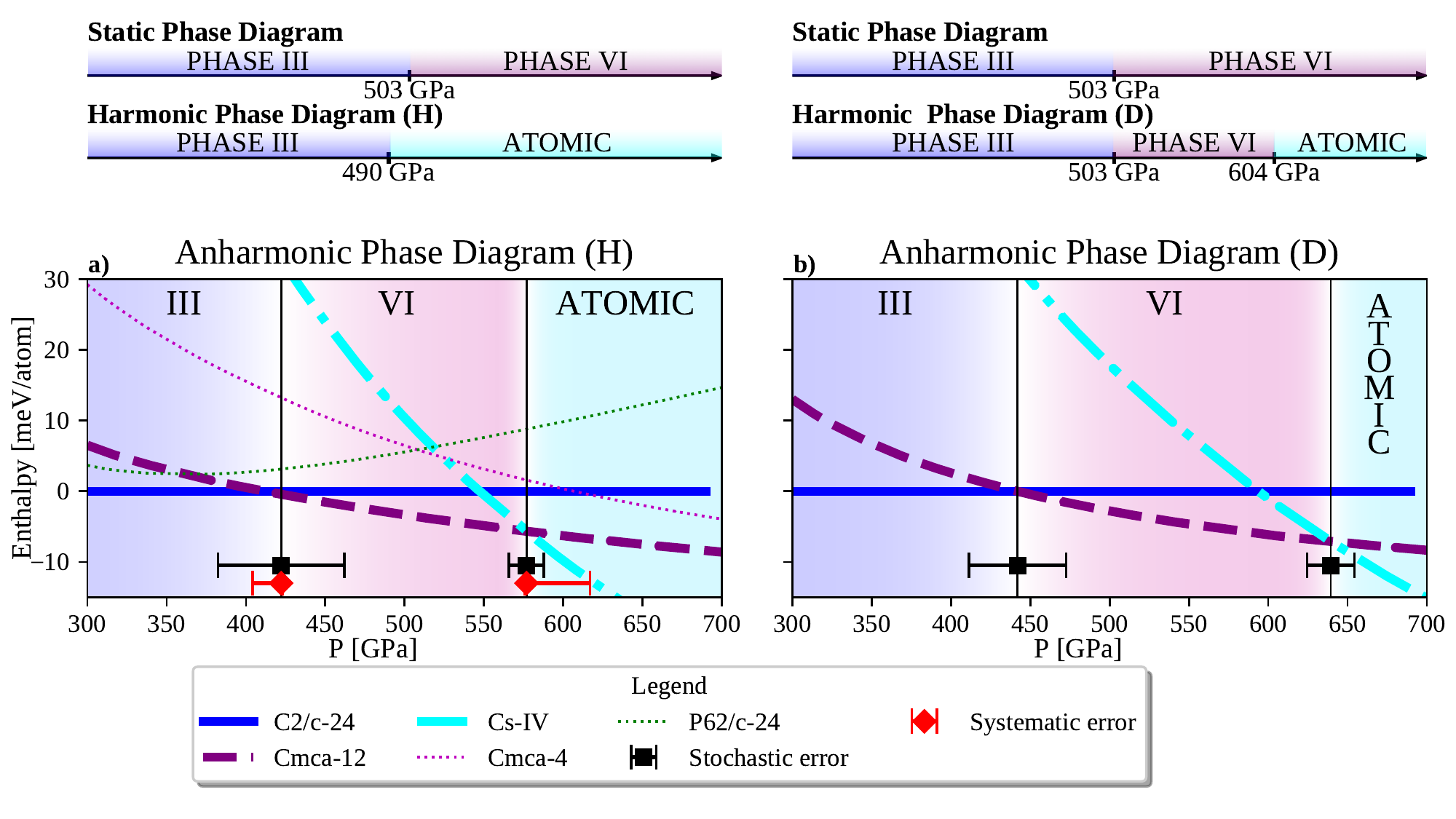}
    \caption{\small
    Phase diagram of hydrogen (H: panel \textbf{a}) and deuterium (D: panel \textbf{b})
    at $T = \SI{0}{\kelvin}$. On the top we report the phase diagram obtained by neglecting nuclear zero-point energy (static lattice) and with the harmonic zero-point energy.
    For the final anharmonic phase diagram we explicitly report the enthalpies of different phase with respect to phase III (C2/c-24). The most stable structure is the lowest in enthalpy at a given pressure.
    Enthalpies are evaluated at the DMC level.
Hydrogen transforms from phase-III (C2/c-24) to phase-VI (Cmca-12) at    \SI{422\pm40}{\giga\pascal} (\SI{442\pm30}{\giga\pascal} for D), and then to the atomic superconductive phase (Cs-IV) at \SI{577\pm10}{\giga\pascal}
  (\SI{640\pm14}{\giga\pascal} for D).  Anharmonicity strongly affects the transition pressures and  qualitatively affects the phase diagram of hydrogen. Stochastic errors (black bars) are linearly propagated from DMC and SSCHA errors in total energy to pressure.  Systematic errors (red bars) are estimated by changing the reference structure for DMC calculation (see SM for more details). 
} 
    \label{fig:PD}
\end{figure*}

We present the complete phase diagram for hydrogen (H) and deuterium
(D) in Figs.~\ref{fig:PD}\textbf{(a)} and \ref{fig:PD}\textbf{(b)}, respectively.
On the top of \figurename~\ref{fig:PD}, we report also the phase diagram computed by neglecting the nuclear zero-point motion (static lattice), and the one with the harmonic zero-point energy.

Hydrogen transforms into the atomic metal 
at \SI{577\pm 15}{\giga\pascal}, much above 
the pressure
predicted neglecting anharmonicity. 
The isotope shift of this transition is 
one of the biggest ever reported, as
deuterium transits into its atomic metallic state at
\SI{640\pm14}{\giga\pascal}. 
Anharmonicity modifies the structure of all molecular phases,
stretching the molecular bonds and softening the \ch{H2} molecular
vibrations 
by about
\SI{1000}{\per\centi\meter}\cite{MonacelliNatPhys2020}. Thus, the
relaxation of anharmonic energy strongly favors molecular phases. 

Even though the anharmonic contributions significantly impact the
energy difference between molecular and atomic phases, the latter
turns out not to be as harmonic as suggested previously\cite{Borinaga2017}. Indeed, we found that Cs-IV exhibits a
prominent anharmonicity in the cell shape.  
The only free parameter of the Cs-IV structure is the $c/a$ ratio of the tetragonal lattice. 
The anharmonicity increases the $c/a$ ratio by about 0.12,
independently of the pressure and the level of the electronic theory
employed.
A recent 
path integral molecular dynamics
calculation also showed a
nontrivial anharmonicity in the Cs-IV phase\cite{Morresi2021}, 
in agreement with us. 
The correct simulation of the $c/a$ structural parameter has relevant consequences on 
the superconducting
properties: by varying the $c/a$ at a fixed volume, the Cs-IV
undergoes a Lifshitz transition that enhances the density of states (DOS) at the Fermi
level. Quantum anharmonic fluctuations shift 
$c/a$ away from the
Lifshitz transition, 
preventing the enhancement of
the superconducting critical
temperature 
in the range of pressure where this phase is
stable (see SM). 

\begin{figure*}
\centering
\includegraphics[width=\textwidth]{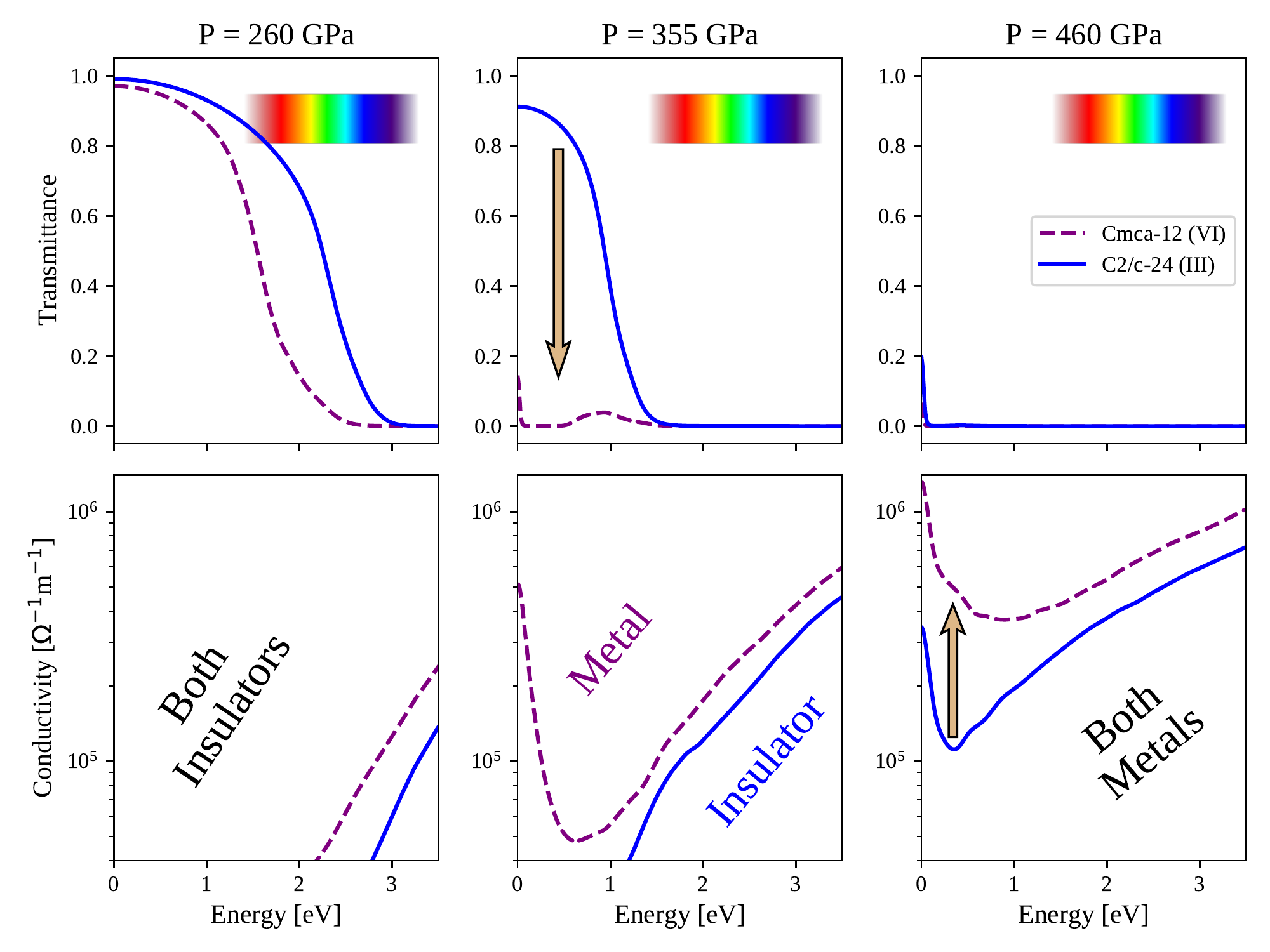}
\caption{\small Simulation of transmittance (upper panels) and
  the real part of optical conductivity (lower panels)
  at various pressures, comparing phase III
  (C2/c-24, blue solid lines) with phase VI (Cmca-12, violet dashed lines). The rainbow colors
  match their respective energy in the visible 
  spectrum. The arrows 
  highlight
  differences in the optical properties between the two phases.
  We observe the transmittance of phase VI dropping down at lower energies than phase III. This phase is already completely opaque at \SI{355}{\giga\pascal}. Also, DC conductivity is higher in phase VI than phase III. The increase of conductivity around zero energy is the Drude peak and it is a signature of the indirect band gap closure (metallicity). Reflectivity and electronic density of states are reported in SM (\figurename~\ref{fig:reflect}).
  } 
\label{fig:optic}
\end{figure*}

\begin{figure}
    \centering
    \includegraphics[width=0.98\textwidth]{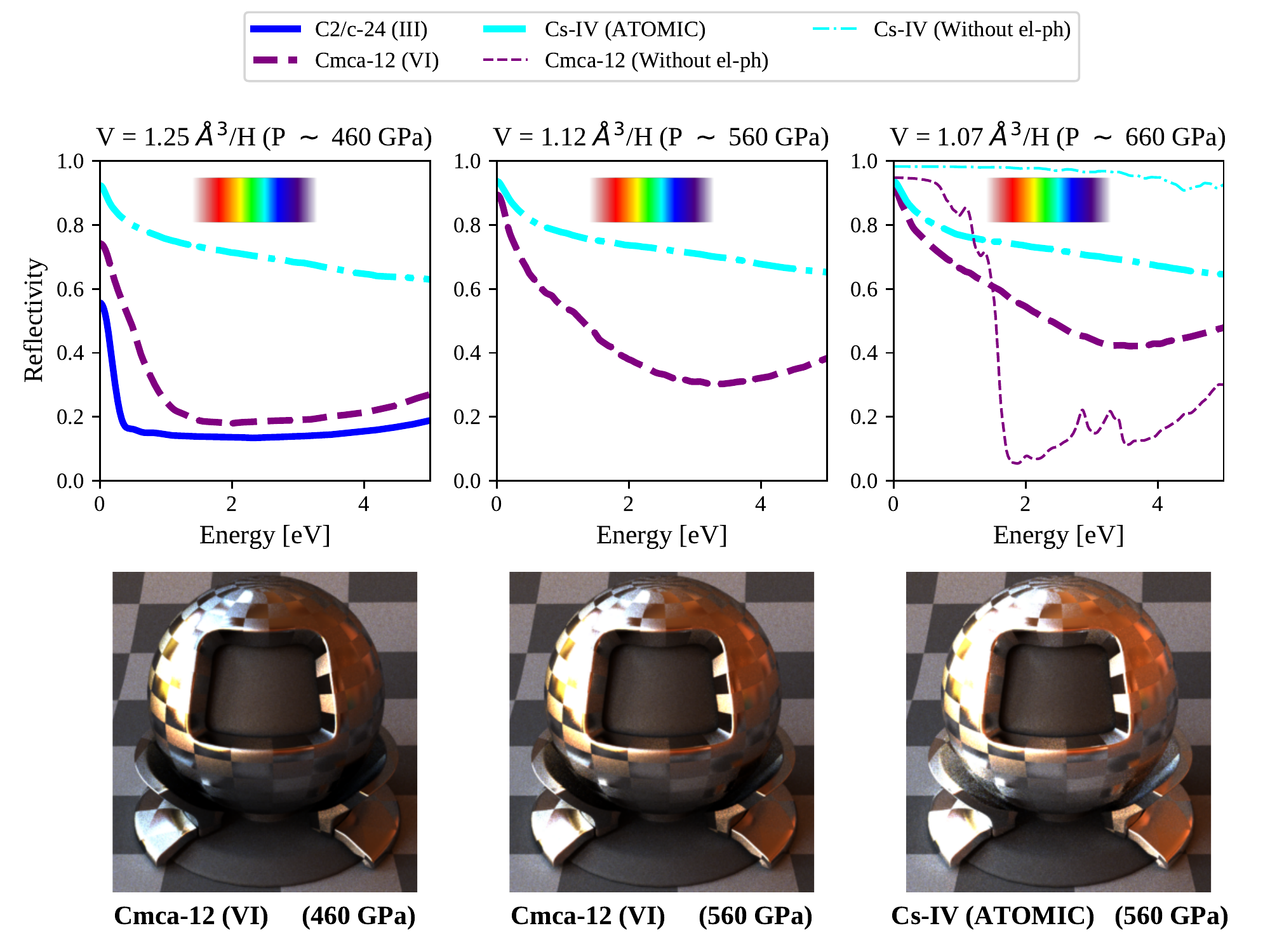}
    \caption{Reflectivity (upper panels)
    and photo-realistic render (lower panels)
    of high-pressure metallic hydrogen. At the transition between phase III to phase VI there is no visible jump in the reflectivity under visible light and the metal appears almost black. As we increase the pressure, the reflectivity gradually increases until we reach the transition to the atomic phase, where we have a significant jump and the material appears shiny (bottom-right panel). In the top-right panel, we also report the reflectivity obtained without electron-phonon (el-ph) coupling (i.e. a standard static calculation). Interestingly, in molecular phases, the electron-phonon coupling enhances the reflectivity by increasing the DOS at the Fermi level, while, for atomic hydrogen, electron-phonon suppresses it. The data are calculated at the same volume for all phases. For the Cs-IV phase, the corresponding pressure is about \SI{40}{\giga\pascal} lower than the one reported in the title (which is approximately the same for C2/c-24 and Cmca-12). The photo-realistic render of the high-pressure hydrogen in vacuum is made by feeding the Mitsuba2\cite{MITSUBA} software with the complex refractive index of hydrogen and introducing a texture to simulate both a rough and a smooth appearance, following the methodology discussed in\cite{Prandini2019}.
}
    \label{fig:reflect:new}
\end{figure}

Before becoming an atomic metal, hydrogen undergoes another phase transition between two molecular phases (III$\rightarrow$ VI). This
transition occurs at \SI{422\pm 40}{\giga\pascal} for H (\SI{442\pm30}{\giga\pascal} for D), in agreement with the experimental results of Loubeyre \emph{et al.}\cite{Loubeyre2019observation}.
The experimental marker of 
this
phase transition is a sudden drop of the transmittance in the near IR region, ascribed to the closure of the direct bandgap in correspondence to 
a
structural rearrangement. However, other experiments exploring the same pressure range observed the sample under visible light without spotting any trace of 
phase transition\cite{Silvera2017,Eremets2019_hydro}.
To 
investigate this situation,
we computed the optical properties in the near IR and visible range for phase III (C2/c-24), and for the structure we predict to be stable above \SI{422}{\giga\pascal}, namely Cmca-12. 
Our calculations account for electron-phonon interaction non-perturbatively. The electronic bands are computed within the modified Becke-Jonson meta-GGA\cite{Tran_2009}, which shares a similar accuracy 
with
hybrid functionals and self-consistent GW calculations, by following the same methodology discussed in \cite{MonacelliNatPhys2020}.
We find that the Cmca-12 structure does not transmit light in the IR around the transition pressure, in contrast with phase III (C2/c-24), as shown in \figurename~\ref{fig:optic}. Our data explain the drop in IR transmittance observed in the experiment\cite{Loubeyre2019observation} and, thus,  support the assignment of phase VI to the Cmca-12 symmetry. 
Moreover,
both phase III and VI display an almost identical low reflectivity in the visible window of 1.8-\SI{3.2}{\electronvolt}  (\figurename~\ref{fig:reflect:new}), so they are almost indistinguishable under visible light. This explains why 
experiments that explored the required pressure did not observe the phase transition\cite{Dias2017,Eremets2019_hydro}.
We also predict a resistance drop 
at
the phase transition, associated with an increase of the electronic DOS at the Fermi level. Conductivity measurements on hydrogen\cite{Eremets2019_hydro} stop just before the
transition pressure. The sudden rise in conductivity is an independent feature that can unambiguously prove the transition to phase VI.
Our results differ from a previous theoretical work\cite{Gorelov2020}, which adopted a different approximation,
where
optical electron-hole excitations involving phonons outside the center of the Brillouin zone are 
not included.
Interestingly, 
Cmca-12 is never the ground state of hydrogen with harmonic zero-point energy, and it is stabilized by anharmonicity.

In \figurename~\ref{fig:reflect:new}, we report the reflectivity data for the phases III, VI and the atomic one as pressure increases. At each pressure, we show a photo-realistic render of high-pressure solid hydrogen sample in vacuum, simulated by solving the Fresnel equation as implemented in the Mitsuba2 software\cite{MITSUBA}.
Phase VI becomes gradually more reflective upon
increasing pressure, until it transforms into the atomic Cs-IV at about \SI{577}{\giga\pascal}, where it becomes shiny, reflecting almost 80 \% of the visible light. Despite being significantly attenuated by vibrational disorder, the sudden rise of reflectivity in the visible light is a key signature of the molecular-to-atomic transition.
Together with reflectivity, also the static conductivity gradually increases upon loading pressure and jumps to higher value at the transition to the atomic Cs-IV phase (see SM).
In contrast to phase VI, the atomic phase shows no significant variation of reflectivity and conductivity with pressure. Interestingly, the quantum nuclear fluctuations have an opposite effect on molecular phases, where they enhance reflectivity, 
than on 
the atomic phase, in which the reflectivity is strongly suppressed. The suppression of reflectivity in the atomic phase was already found in other works\cite{Borinaga_2018,Zhang_2018}.

In conclusion, the hydrogen phase diagram based on both highly
accurate electronic internal energies computed by QMC and anharmonic
nuclear quantum fluctuations provided by SSCHA, 
confirms that hydrogen undergoes a first-order
phase transition from a conductive phase III (molecular C2/c-24) to metallic phase VI (molecular Cmca-12) at
\SI{422\pm40}{\giga\pascal}, in accordance with experiments\cite{Loubeyre2019observation}. 
We predict the transition towards atomic metallic hydrogen 
to occur at about \SI{577}{\giga\pascal}, a much larger pressure than those reached so far by experiments\cite{Dias2017}.
Our results indicate
that the synthesis of atomic hydrogen has still to be fulfilled.

\bibliography{biblio}

\bibliographystyle{Science}

\section*{Acknowledgments}
L.M. acknowledges CINECA under the ISCRA initiative for providing high performance computational resources employed in this work.
 M.C. thanks GENCI for providing computational resources under the grant number 0906493, the Grands Challenge DARI for allowing calculations on the Joliot-Curie Rome HPC cluster under the project number gch0420.
 M.C., K.N., and S.S. thank RIKEN for providing computational resources of the supercomputer Fugaku through the HPCI System Research Project (Project ID: hp210038).
 K.N. acknowledges support from the JSPS Overseas Research Fellowships, from Grant-in-Aid for Early Career Scientists (Grant No.~JP21K17752), and from Grant-in-Aid for Scientific Research (Grant No.~JP21K03400).
 S.S. acknowledges support from MIUR, PRIN-2017BZPKSZ.
 This  work  was supported  by  the  European  Centre of Excellence in Exascale Computing TREX-Targeting Real Chemical Accuracy at the Exascale. This project has received funding from the European Union’s Horizon 2020 Research and Innovation program under Grant Agreement No.~952165.


\section*{Supplementary materials}

\paragraph*{Details on the Phase Diagram calculation and the role of anharmonicity and electronic correlations}

In this Section, we describe in details how the phase diagram is computed, and discuss how it changes when we adopt a different electronic theory - density functional theory (DFT) versus diffusion quantum Monte Carlo (DMC) -  with and without anharmonicity.

We relaxed each structure by including quantum fluctuations and anharmonicity 
through the stochastic self-consistent harmonic approximation (SSCHA), optimizing the auxiliary force constants, centroid positions, and lattice vectors 
within the constraints of the symmetry group, at roughly every 100 GPa (from \SI{250}{\giga\pascal} to \SI{650}{\giga\pascal}).
In the SSCHA calculations, we employed the DFT framework
with the BLYP\cite{BLYP} exchange-correlation functional to account for electronic energy and determine the Born-Oppenheimer (BO) potential energy surface. 
BLYP is one of the most accurate DFT functionals for phase-diagram calculations 
of
high-pressure hydrogen, 
outperforming more refined techniques such as hybrid DFT\cite{Drummond2015,Clay_2014}.
The full anharmonic energy is obtained within DFT, by fitting with a parabola the difference between the BO energy and the SSCHA total energy at fixed volume for each phase. Also the anharmonic stress tensor is employed in the fit to increase 
accuracy. We then add to the static BO energy-versus-volume curves,
computed in DFT every \SI{5}{\giga\pascal}, the quantum anharmonic lattice vibrational contribution at the corresponding volume calculated from the fit. We finally perform the Legendre transform to get the enthalpy-vs-pressure curves and the resulting phase diagram.

\begin{figure}[b!]
    \centering
    \includegraphics[width=.8\textwidth]{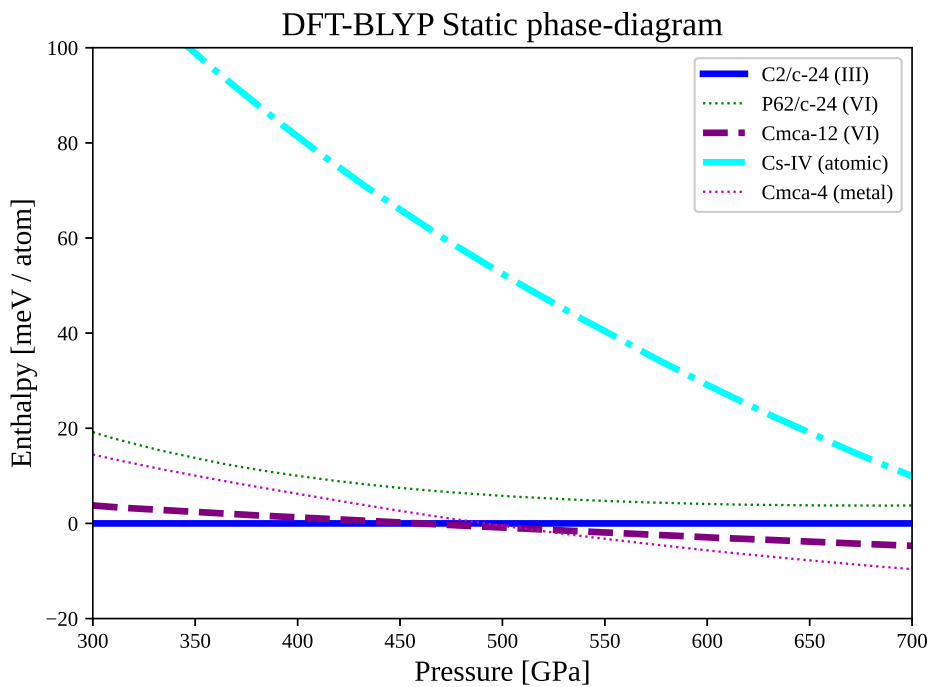}
    \caption{DFT-BLYP static enthalpies 
    (static lattice)
    of high-pressure hydrogen.}
    \label{fig:staticDFT}
\end{figure}

\begin{figure}
    \centering
    \includegraphics[width=.7\textwidth]{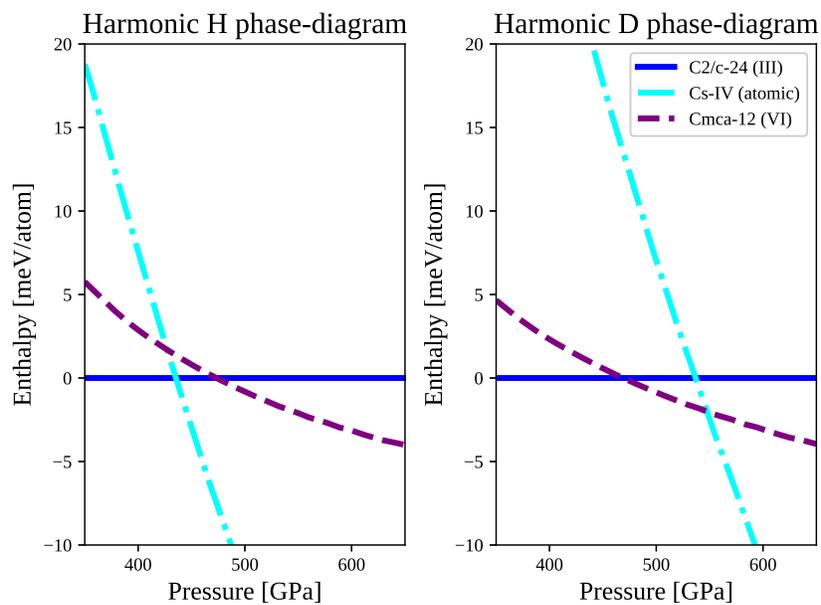}
    \caption{DFT-BLYP enthalpies including nuclear zero-point energy within the harmonic approximation for hydrogen and deuterium, shown on the left and right side, respectively.}
    \label{fig:pd:qha:blyp}
\end{figure}

\begin{figure}
    \centering
    \includegraphics[width=.7\textwidth]{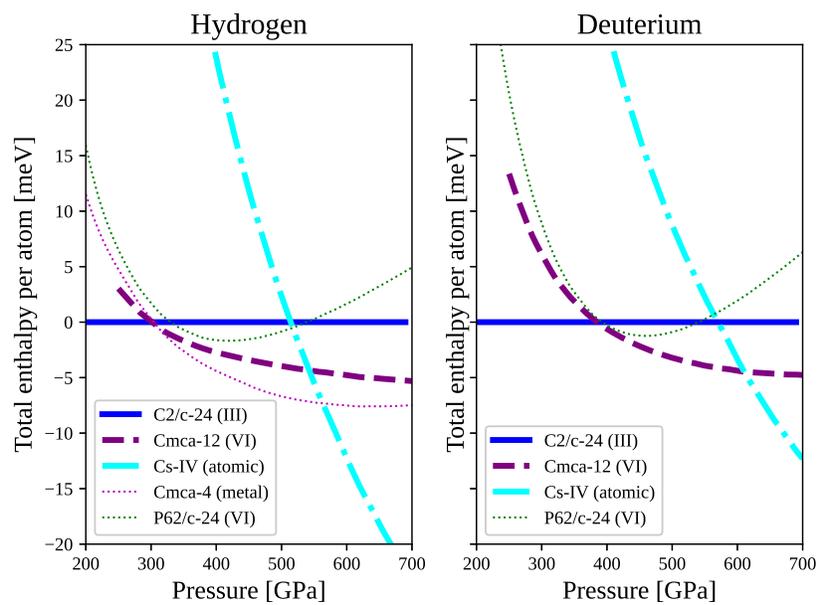}
    \caption{As in Fig.~\ref{fig:pd:qha:blyp}, but for the
    DFT-BLYP enthalpies including quantum anharmonic effects.
}
    \label{fig:pd:sscha}
\end{figure}

The static phase diagram simulated within DFT-BLYP is reported in \figurename~\ref{fig:staticDFT}, while in \figurename~\ref{fig:pd:qha:blyp} we show the DFT-BLYP phase diagram with \emph{harmonic} zero-point energy . We included the harmonic contributions only for the most relevant phases: C2/c-24, Cmca-12 and Cs-IV. 
The harmonic zero-point energy leaves almost unchanged the pressure of the C2/c-24 to Cmca-12 transition (phase III to VI), while it substantially shifts 
the atomic transition 
down
to pressures even lower than the Cmca-12. 
The results of the anharmonic phase diagram of both hydrogen ($^1\text{H}$ protium or H) and deuterium ($^2\text{H}$ or D) computed 
by DFT-BLYP and
SSCHA are reported in \figurename~\ref{fig:pd:sscha}. 
It shows that
anharmonicity strongly favors the molecular phases over the atomic one, shifting back the atomic transition to higher pressures. Between 
Cmca-12 and C2/c-24, 
anharmonicity favors the Cmca-12 crystal symmetry, 
moving the III-to-VI phase transition down by about \SI{150}{\giga\pascal}. In this case, phase VI 
candidates
P62/c-24, Cmca-12 and Cmca-4 
are almost degenerate
up to \SI{400}{\giga\pascal}, where the Cmca-4 starts dominating over the other molecular phases.

Apart from Cmca-4, the DFT-BLYP phase diagram is in qualitative agreement if compared to the one 
including the electron correlation treated at the quantum Monte Carlo (QMC) level.

\newpage
Thanks to extensive DMC calculations performed at fixed structures for several phases and volumes, we have been able to correct the DFT-BLYP internal energies, and add the contribution coming from a nearly exact treatment of electron correlation on the top of the static, harmonic and 
quantum
anharmonic phase diagrams previously computed at the DFT-BLYP level. DMC corrections are added on the total energy-versus-volume curves of the corresponding DFT (and DFT+SSCHA) calculations. As in the DFT case, the enthalpy-versus-pressure curves are then obtained by Legendre transform.

\begin{figure}[b!]
    \centering
    \includegraphics[width=.7\textwidth]{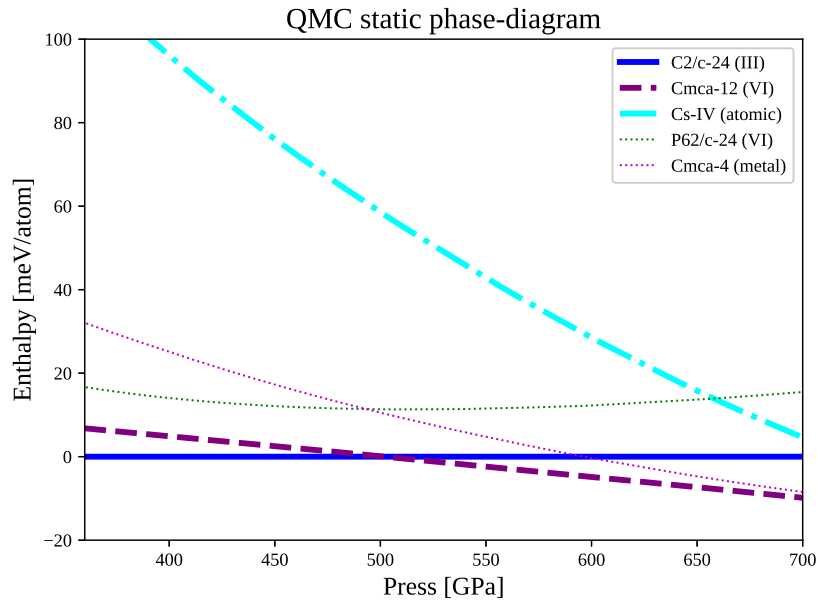}
    \caption{Diffusion QMC static enthalpies 
    (static lattic)
    of high-pressure hydrogen. Based on these enthalpies, we draw the static-nuclei phase diagram reported in the main text (\figurename~\ref{fig:PD}).
    }
    \label{fig:qmcstatich}
\end{figure}

\begin{figure}
    \centering
    \includegraphics[width=.65\textwidth]{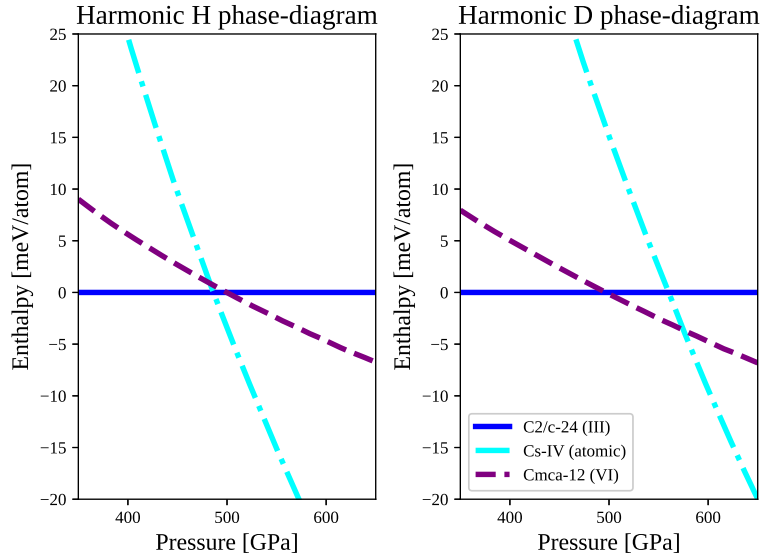}
    \caption{Diffusion QMC enthalpies including nuclear zero-point energy within the harmonic approximation for hydrogen (left panel) and deuterium (right panel). 
    Based on these enthalpies, we draw the harmonic phase diagram reported in the main text (\figurename~\ref{fig:PD}).
    The harmonic zero-point energies are calculated at the DFT-BLYP level and then added to the DMC energies.
    }
    \label{fig:QHA}
\end{figure}

\begin{figure}
    \centering
    \includegraphics[width=.65\textwidth]{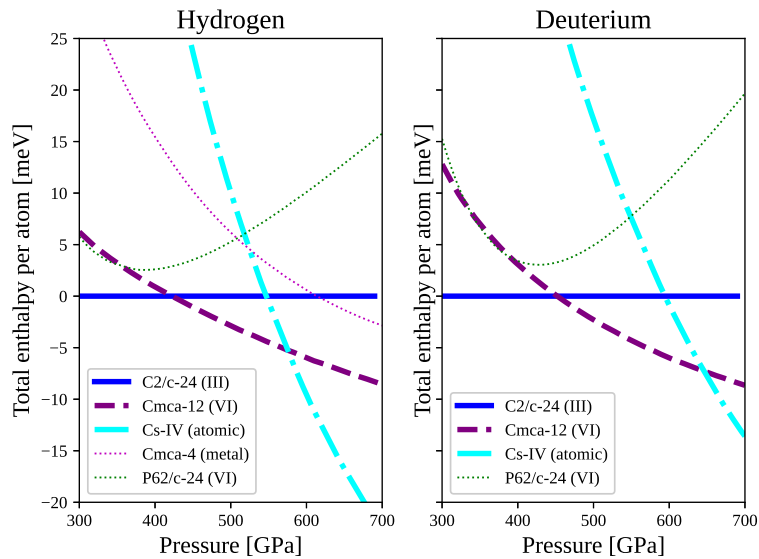}
    \caption{Diffusion QMC enthalpies, including quantum anharmonic effects for hydrogen and deuterium, shown on the left and right side, respectively. Nuclear quantum effects are added based on SSCHA calculations performed at the DFT-BLYP level. The corresponding phase diagram is reported in the main text (\figurename~\ref{fig:PD}).
    }
    \label{fig:pd:full}
\end{figure}

For the sake of completeness, we report the static DMC-corrected phase-diagram in \figurename~\ref{fig:qmcstatich}, the DMC-corrected enthalpies accounting for the nuclear zero-point energy within the harmonic approximation (\figurename~\ref{fig:QHA}) and the full anharmonic enthalpies with DMC corrections (\figurename~\ref{fig:pd:full}). The latter data provide the final phase diagram reported in the main text.

\newpage

\paragraph*{The P62/c-24 symmetry}

Among other known structures, we also simulated the new P62/c-24 symmetry 
we discovered through the relaxation of phase III (C2/c-24) at the anharmonic level within DFT-BLYP.

\begin{figure}[b!]
    \centering
    \includegraphics[width=0.7\textwidth]{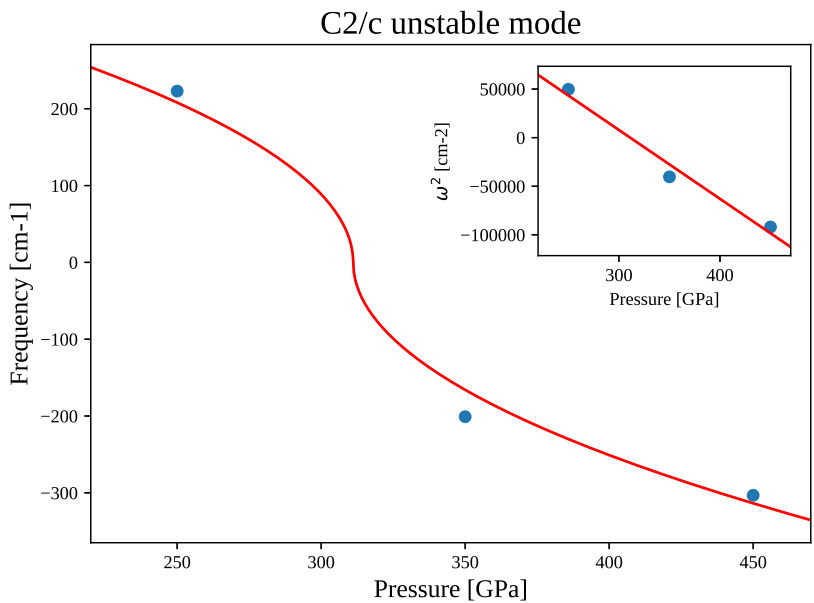}
    \caption{Frequency of the eigenvalue of the free energy Hessian along the unstable nuclear vibration, obtained with DFT-BLYP. On the negative axis we report imaginary values. Inset: the square of the frequency, which correspond to the free energy curvature. A negative value indicates an instability. 
    }
    \label{fig:c2ctop62c}
\end{figure}

In particular, phase III becomes unstable at DFT-BLYP level after \SI{310}{\giga\pascal},
when
the free energy curvature becomes negative around an IR-active nuclear vibration at $\Gamma$.
In \figurename~\ref{fig:c2ctop62c} we report the simulation of the C2/c-24 free energy Hessian 
as a function of pressure along with the unstable nuclear vibration. The free energy Hessian at the SSCHA level is computed with the full expression discussed in Ref.\cite{Bianco2017}, including non perturbatively both three- and four-phonon scattering vertices.
Interestingly, this is a very peculiar case where the four-phonon scattering is fundamental to have a correct result, even at a qualitative level, as the C2/c-24 is unstable at all pressures if only three-phonon scattering processes are accounted for.

The unstable mode breaks the C2/c symmetry 
in a Cc group with just two 
symmetry operations
and 24 atoms in the unit cell.
We performed the full anharmonic relaxation of the new phase. The monoclinic cell 
becomes hexagonal, and two layers out of four in the primitive cell 
transform in perfectly graphene-like sheets,
with alternated stacking.
The other two layers keep their molecular feature, and the \ch{H2} molecule reduces its bond length with respect to the C2/c geometry. The transformation of the graphene-like layer is reported in \figurename~\ref{fig:all_structures}b. The symmetry group of the new structure is P62/c, as identified through both ISOTROPY\cite{ISOTROPY} and spglib\cite{SPGLIB} software. This phase is strongly unstable at harmonic level (it has four degenerate imaginary frequencies at $\Gamma$ above \SI{2000i}{\per\centi\meter}) but it is stabilized by anharmonicity.
As far as we know, this is the first example of a new structure discovered by a full quantum relaxation of nuclear position. This is only possible thanks to the simultaneous relaxation of auxiliary force constants, centroids, and lattice vectors.

When more accurate DMC calculations are employed to evaluate its energy, this phase becomes unfavoured. Therefore, 
the instability of 
C2/c-24 towards P62/c is an artifact of the DFT-BLYP functional. 

\paragraph*{The atomic phase}

In the atomic phase,
the only free parameter 
is the c/a ratio of the 
primitive
lattice vectors. 
In the following, we will then present its main properties as a function of the c/a value.

\begin{figure}
    \centering
    \includegraphics[width=0.9\textwidth]{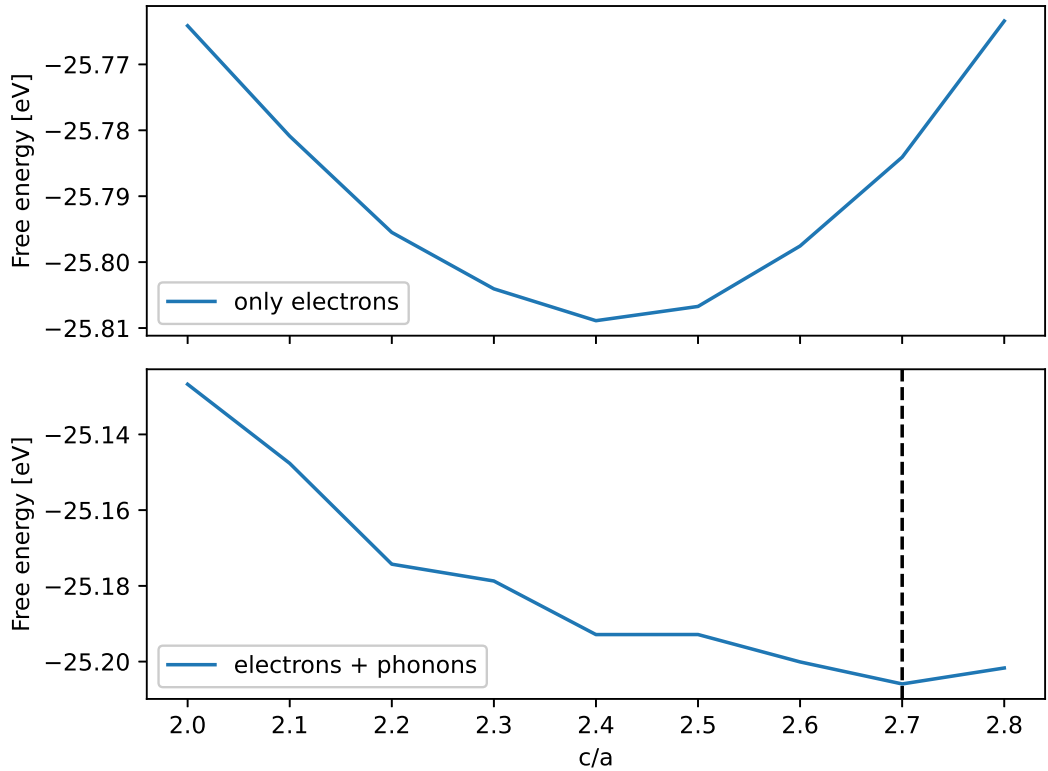}
    \caption{Free energy with static lattice (upper panel) and with harmonic lattice vibrations (lower panel) as a function of c/a. These quantities are computed with the BLYP functional at the volume of $\SI{1.067}{\angstrom}^3$ per atom. The vertical dashed line indicates the c/a value where imaginary phonons appear.}
    \label{fig:qha:ca}
\end{figure}

The structure is stable at the static level, as it has a well-defined minimum. However, suppose we compute phonons at the harmonic level, and use the phonon dispersion to include the kinetic energy of ions due to the quantum zero-point motion. In that case, the total energy decreases with the c/a ratio
until imaginary frequencies appear 
before the minimum is reached, 
and the system becomes unstable (see Fig.~\ref{fig:qha:ca}). The Cs-IV atomic phase is, therefore, unstable within the quasi-harmonic approximation. The SSCHA fixes this instability: the c/a 
increases 
only
by about 0.2 compared to the static value. This effect is, however, strongly 
size-dependent,
and its value becomes even smaller (c/a $\approx$ 0.12) when 
larger cells of 128 atoms are considered, by strengthening the outcome of our analysis.

We computed the free energy Hessian at the SSCHA-relaxed c/a value, and it is stable by a significant amount, 
shifting the higher energy modes down 
only by
about $\SI{250}{\per\centi\meter}$.
Therefore, even if the structure itself is not so anharmonic, the stability of the structure is met only within a complete anharmonic calculation.

It turns out that
the c/a equilibrium value 
is strongly functional dependent. 
Nevertheless,
by running a SSCHA simulation either with BLYP or with PBE, we verified that the effect of the SSCHA on the c/a parameter is additive on the functional used and, thus, the shift with respect to the static equilibrium value is rather functional independent.


\begin{figure}[b!]
    \centering
    \includegraphics[width=0.49\textwidth]{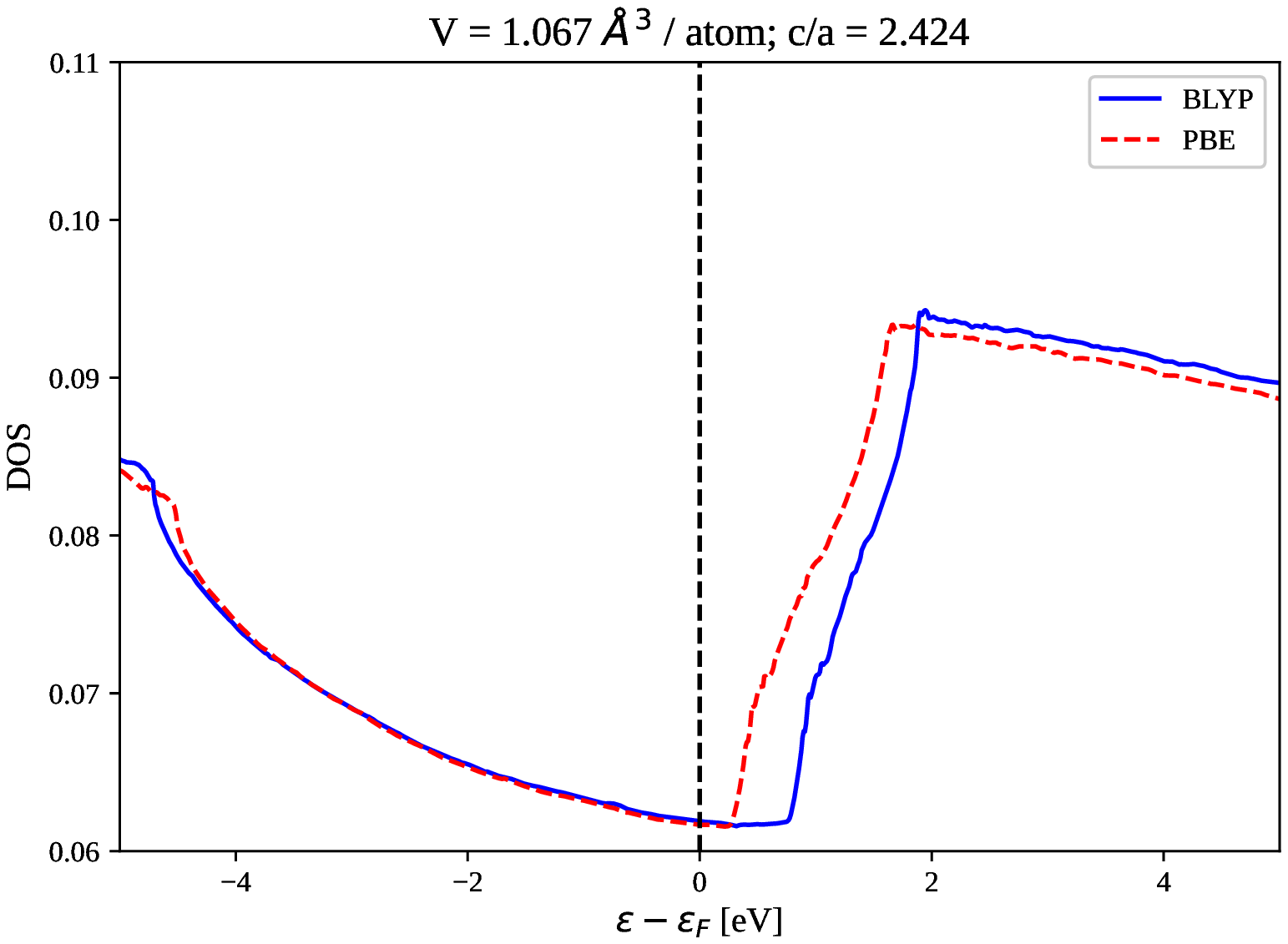}
    \includegraphics[width=0.49\textwidth]{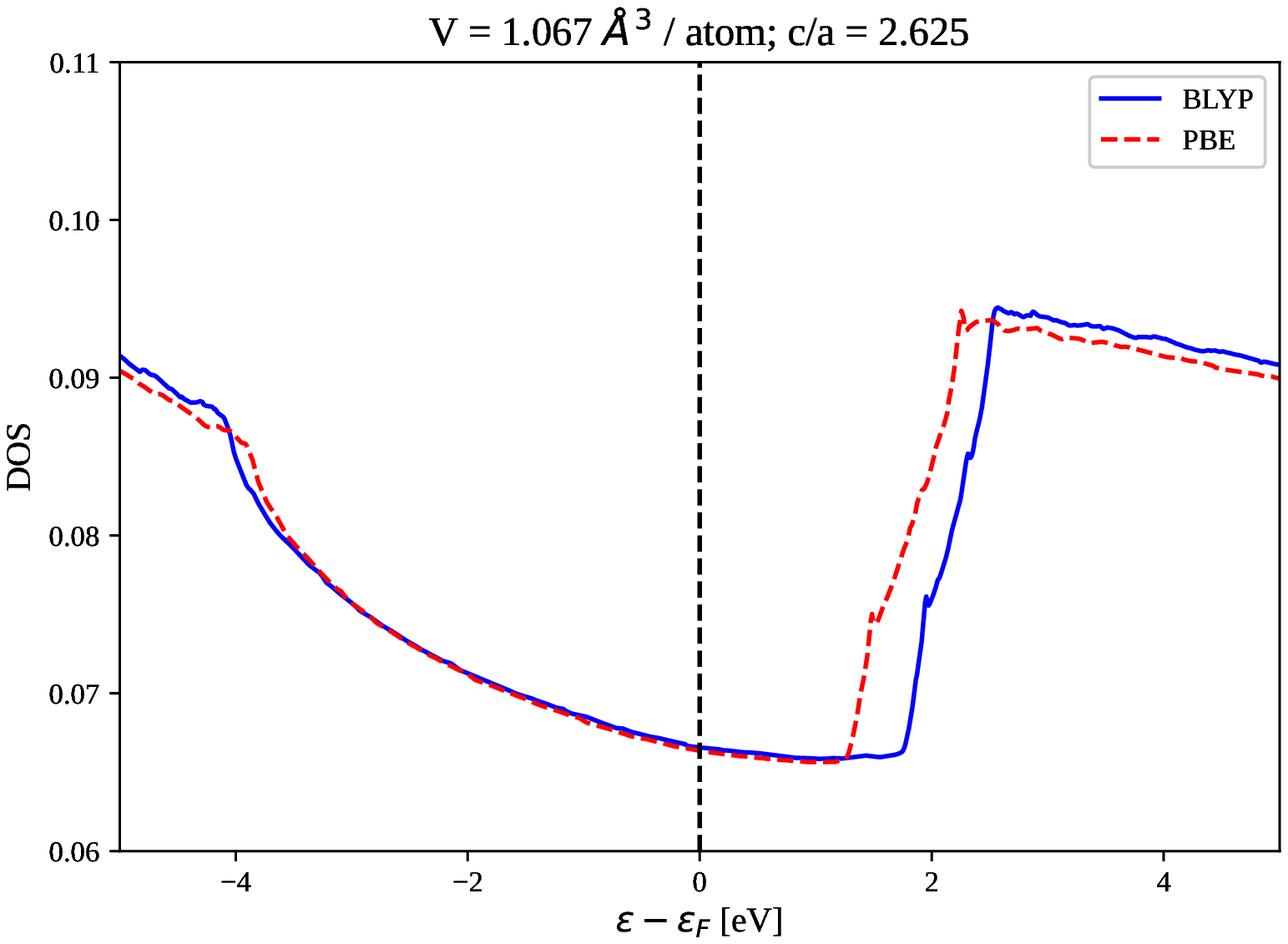}
    \caption{Electronic DOS for c/a 
    values corresponding to BLYP
    static equilibrium geometry (left panel) and 
    to 
    BLYP equilibrium geometry including also nuclear fluctuations (right panel).
    The DOS increases when increasing c/a. However, we get further away from the Lifshitz transition for larger c/a values, and this happens for both BLYP (blue lines) and PBE (red lines) functionals.}
    \label{fig:lifshifts}
\end{figure}

\begin{figure}
    \centering
    \includegraphics[width=0.49\textwidth]{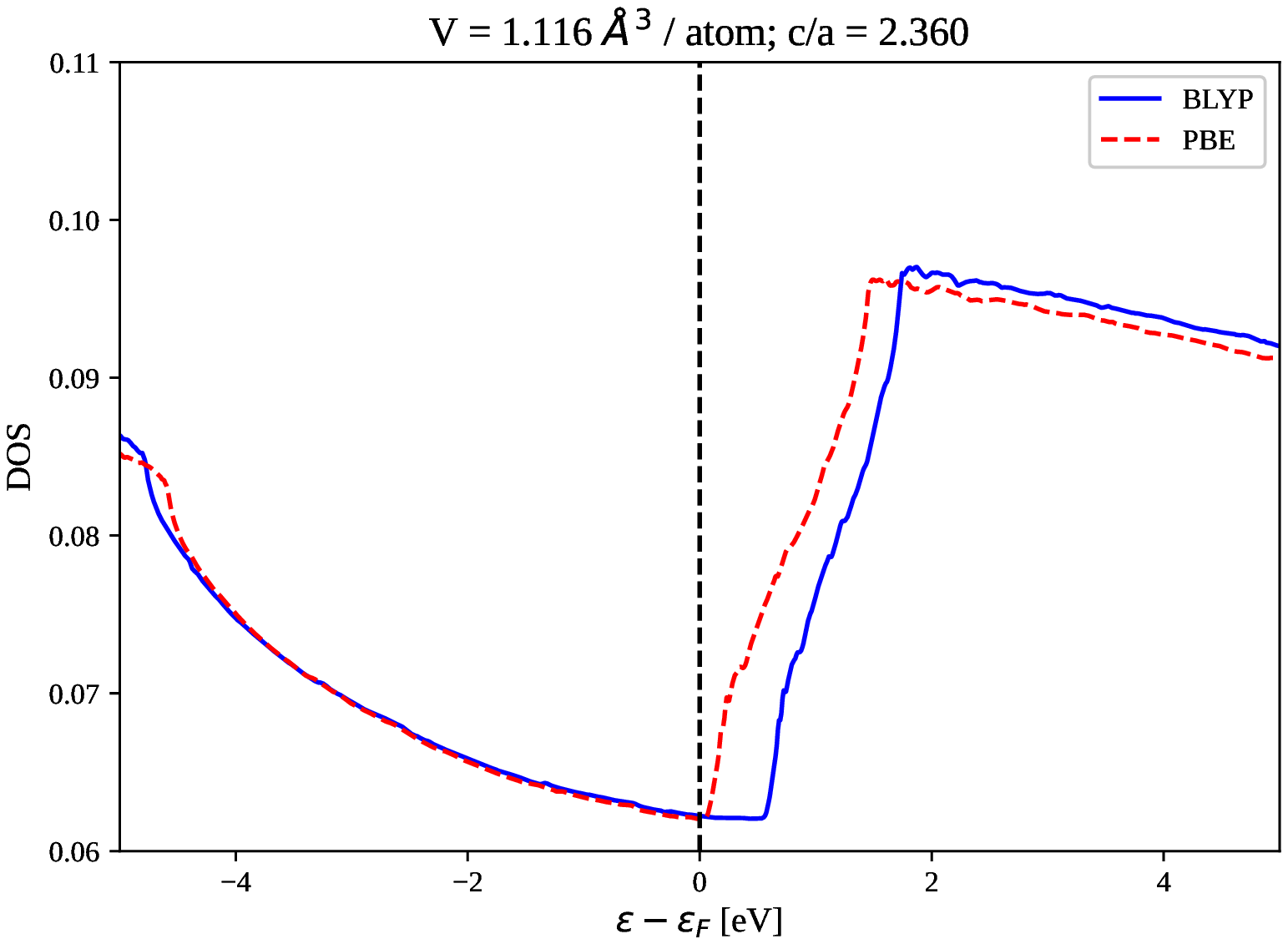}
    \includegraphics[width=0.49\textwidth]{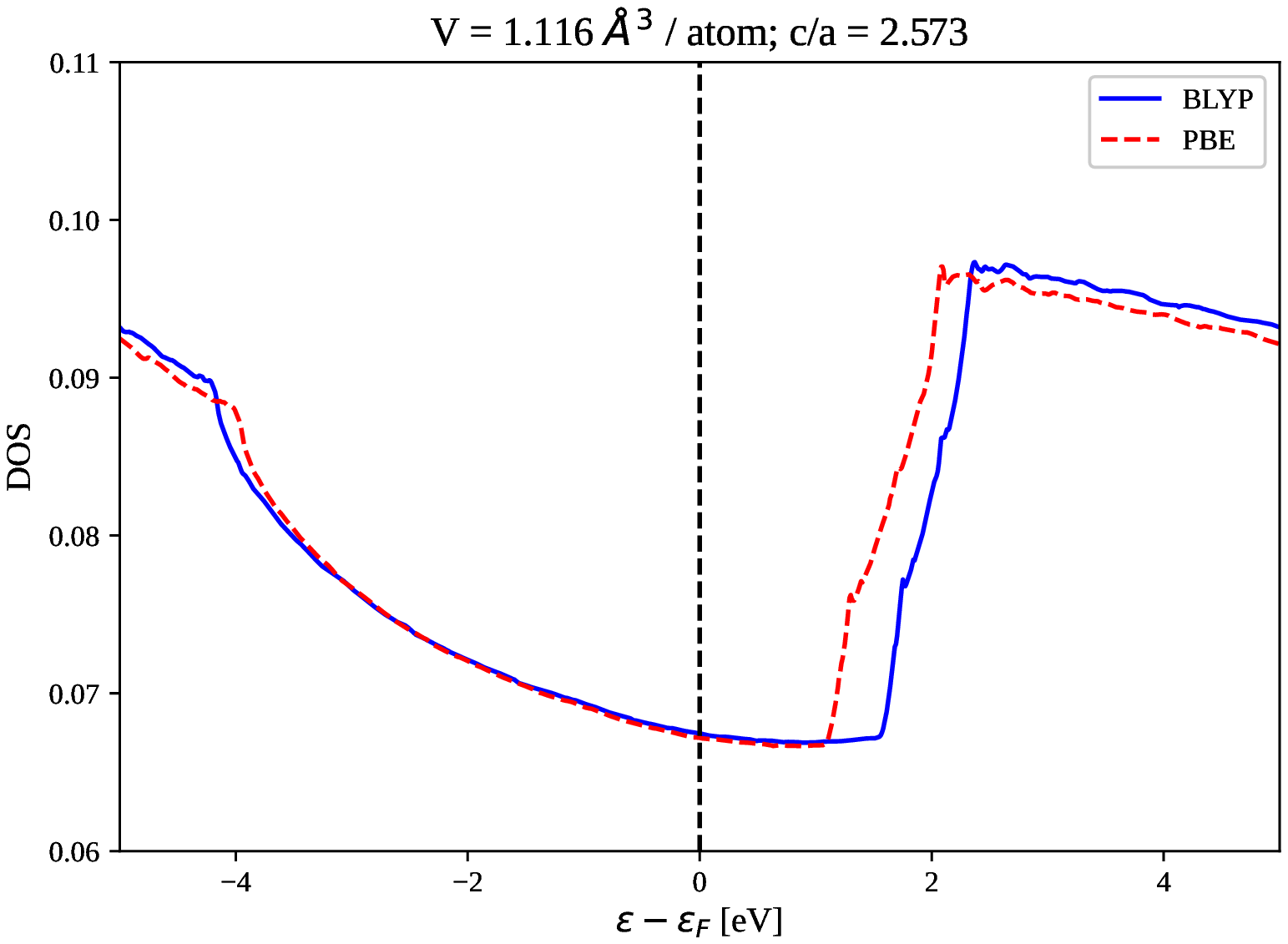}
    \includegraphics[width=0.49\textwidth]{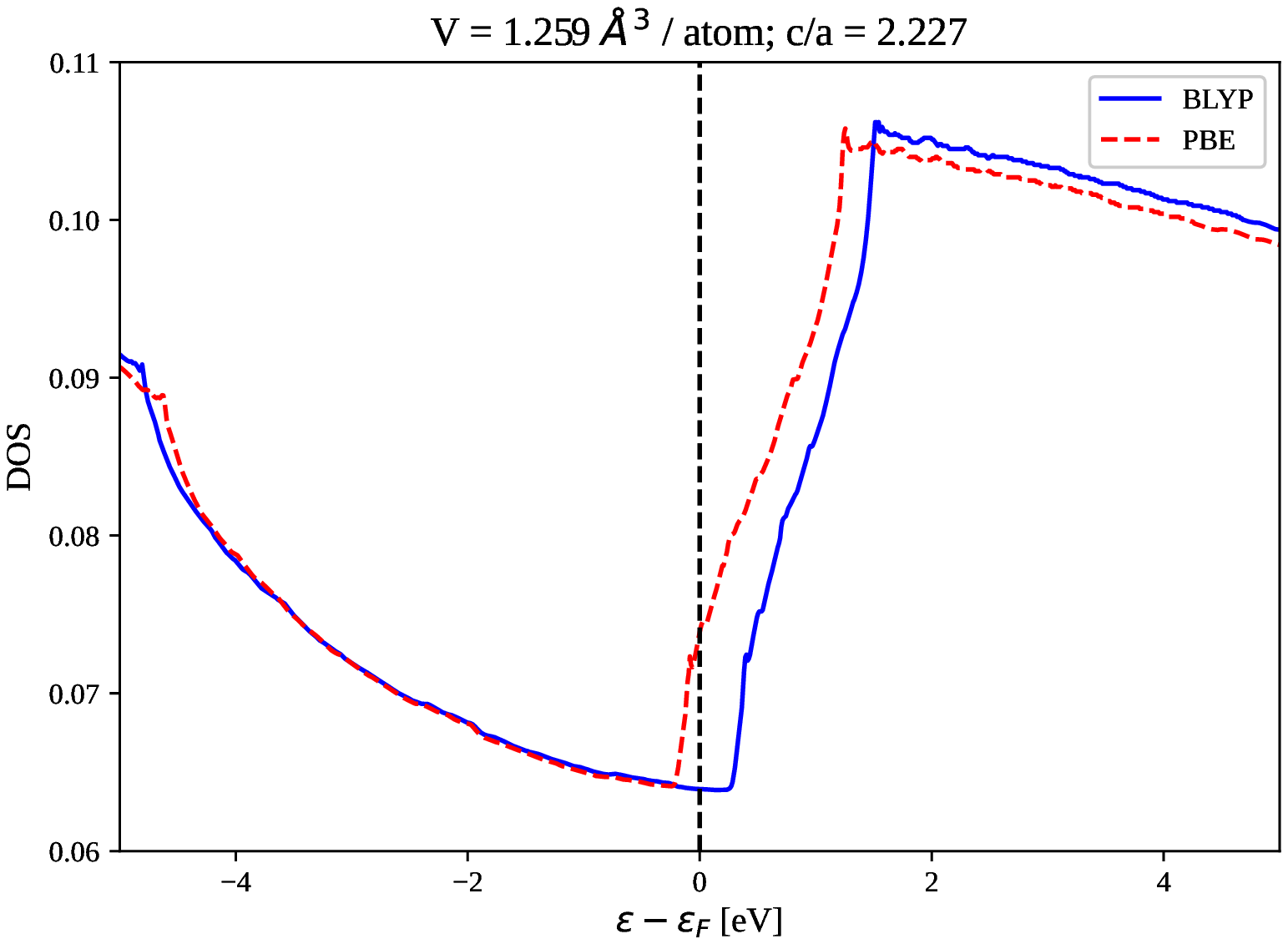}
    \includegraphics[width=0.49\textwidth]{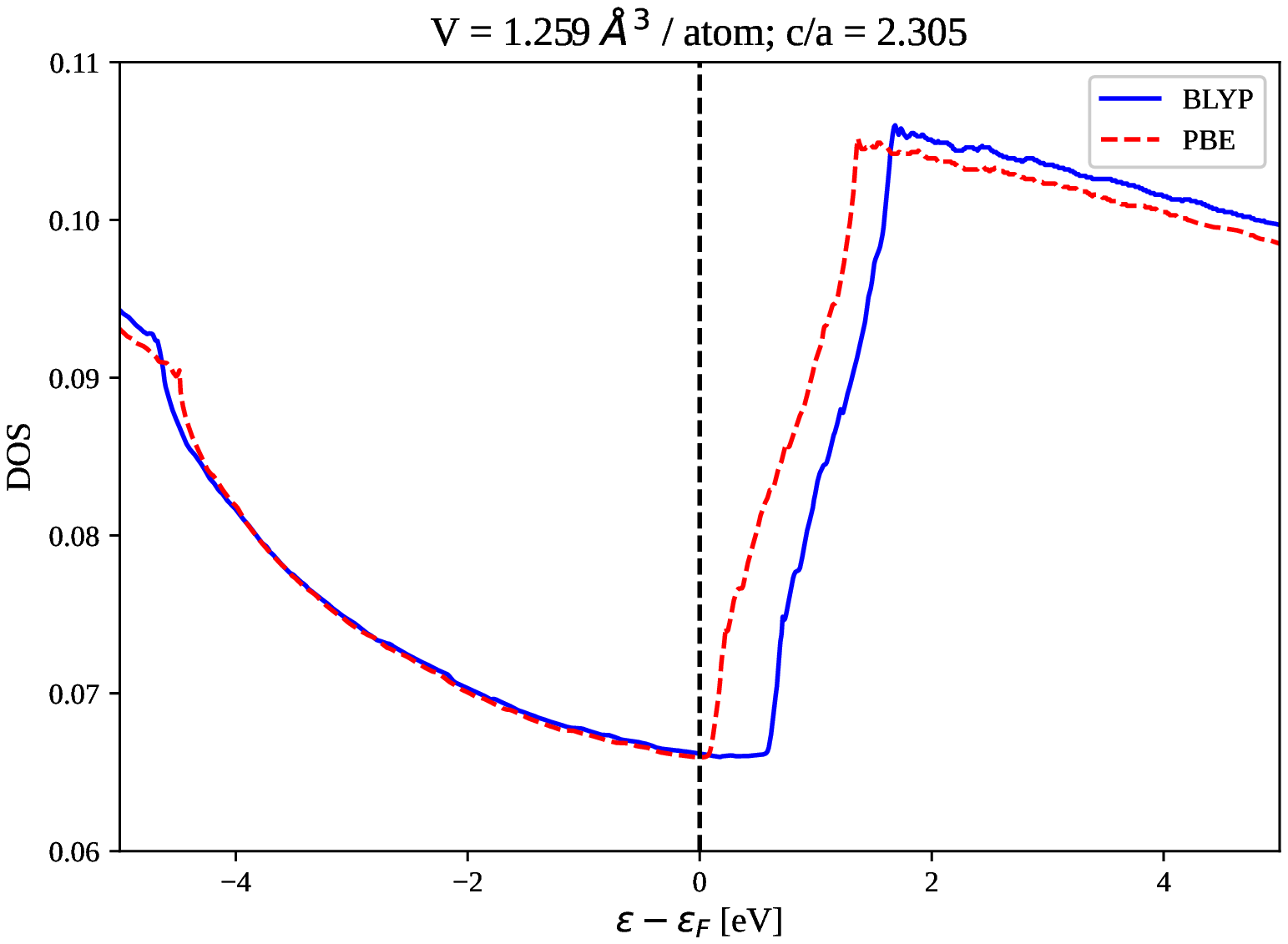}
    \caption{DOS 
     plotted for different volumes and c/a values corresponding to the BLYP static equilibrium geometry (left-hand side), and the BLYP equilibrium geometry with nuclear fluctuations (right-hand side). For these geometries, the DOS is computed with both BLYP (blue lines) and PBE (red lines) functionals.
    }
    \label{fig:lifshifts:v}
\end{figure}

Interestingly, the atomic phase is 
in the proximity of
a Lifshitz transition, signaled by a sudden jump of the 
DOS,
which is indeed located very close to the Fermi level ($\epsilon_F$). This is reported in \figurename~\ref{fig:lifshifts}. In particular, although the DOS at $\epsilon_F$ steadily increases with c/a, 
the Lifshitz transition
gets further away from $\epsilon_F$.
The PBE and BLYP functionals predict the same electronic DOS at $\epsilon_F$ but slightly different locations for the Lifshitz transition, whose energy is systematically closer to the Fermi level in PBE. The same behavior is found also for the other volumes studied for the atomic phase, as reported in \figurename~\ref{fig:lifshifts:v}.
Only at the largest volume taken into account, i.e. V=1.259 \AA$^3$/atom,
the PBE functional triggers the Lifshitz transition, as shown in the bottom-left panel of \figurename~\ref{fig:lifshifts:v}. However, the c/a value reported there corresponds to the BLYP equilibrium geometry 
of the static lattice.
The PBE equilibrium geometry has 
a larger c/a value, which 
pushes the Lifshitz transition energy above the Fermi level.

\begin{figure}
    \centering
    \includegraphics[width=0.49\textwidth]{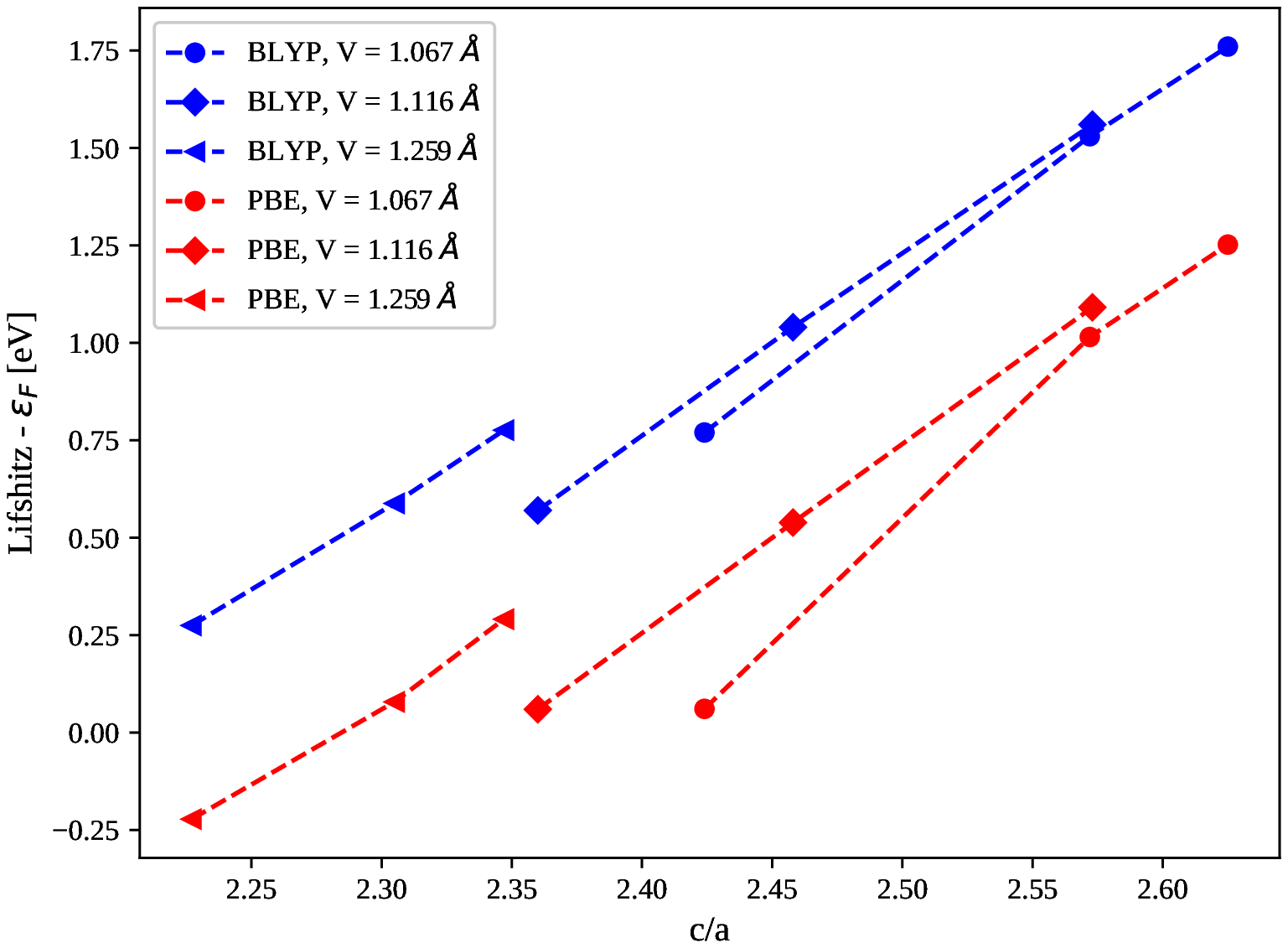}
    \includegraphics[width=0.49\textwidth]{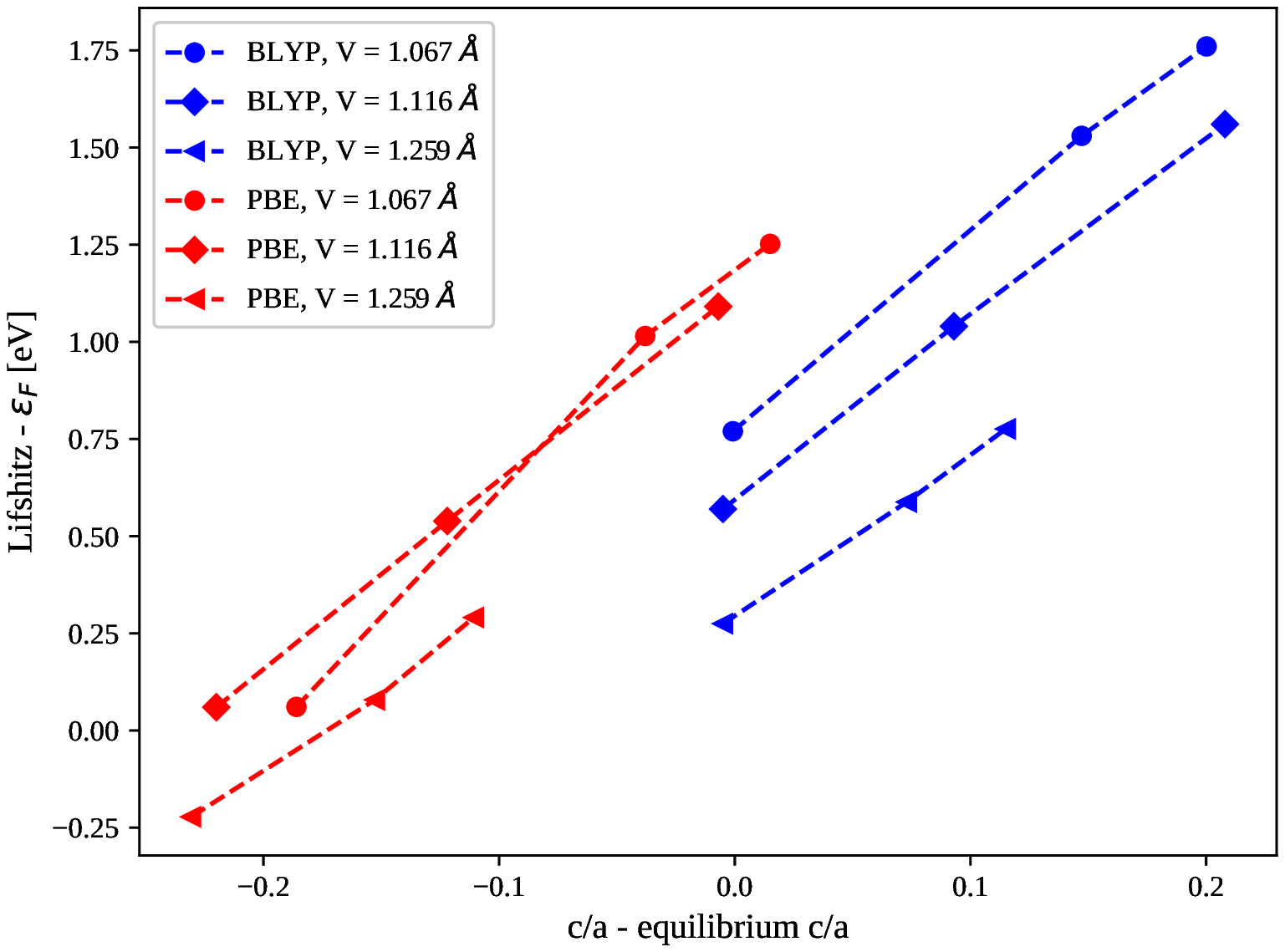}
    \caption{Distance of the Lifshitz transition from the Fermi level.
    In the left panel, this is plotted as a function of c/a. On the right panel, the origin of the c/a axis is set to the equilibrium value of c/a for each volume and functional. The abscissa at which the curves cross the $y = 0$ line is the c/a value when the system undergoes the Lifshitz transition. This is neither volume nor functional universal. 
    }
    \label{fig:lif:all}
\end{figure}

\begin{figure}
    \centering
    \includegraphics[width=.56\textwidth]{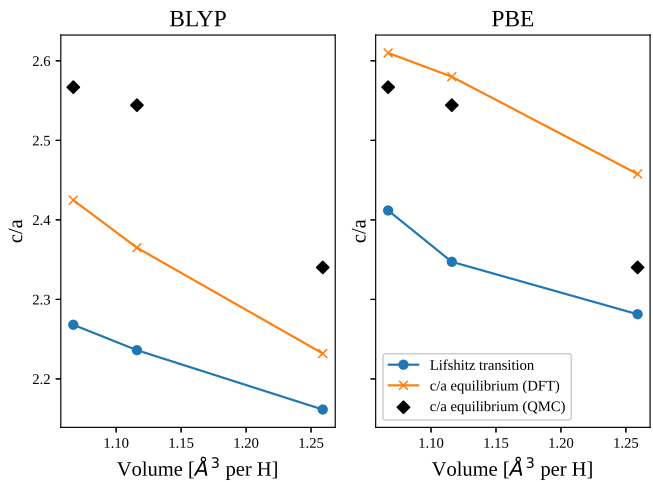}
    \caption{The c/a value at which the Lifshitz transition occurs, causing a sudden increase of the electronic DOS at the Fermi level. We compare the Lifshitz transition with the equilibrium c/a value as yielded by 
    BLYP, 
    PBE, and DMC. 
    Quantum fluctuations move the equilibrium c/a ratio by about 0.1-0.2 towards larger values, independently on the choice of the DFT functional, driving the system away from the transition.}
    \label{fig:lif}
\end{figure}

There is a strong linear
correlation between the energy location of the Lifshitz transition and the c/a value, as revealed by plotting
the distance of the Lifshitz transition energy from $\epsilon_F$
as a function of c/a for all volumes and functionals taken into account (\figurename~\ref{fig:lif:all}).
This correlation allows us to estimate 
the exact c/a value at which the Lifshitz transition occurs for each volume, as obtained by fitting the results in \figurename~\ref{fig:lif:all} and extrapolating the Lifshitz transition energy
to $\epsilon_F$ (\figurename~\ref{fig:lif}).


Care must be taken in evaluating the distance of the Lifshitz transition energy from the Fermi level. Indeed,
the Fermi level determination could be affected by the choice of the smearing parameter. 
In our analysis, 
we employed the so-called "cold" smearing (aka the Marzari-Vanderbilt scheme).
We repeated the calculation for three volumes 
by using the
Gaussian smearing 
and the 
Fermi-Dirac smearing, 
and compared their outcome for the Fermi energy determination.
It turns out
that the difference in the Fermi level is at most \SI{20}{\milli\electronvolt}. The data are reported in \tablename~\ref{tab:smearing}. 
This Fermi energy uncertainty converts into an error on the c/a value for the Lifshitz transition of 0.008, much smaller than the variations plotted in Fig.~\ref{fig:lif}.

\begin{table}[hbtp]
    \centering
    \begin{tabular}{c|ccc}
         \textbf{Volume per \ch{H}} &  \textbf{ Marzari-Vanderbilt} & \textbf{ Gaussian} & \textbf{ Fermi-Dirac}\\
         \hline
         \SI{1.067}{\angstrom^3} &
         \SI{17.115}{\electronvolt} & 
         \SI{17.161}{\electronvolt} & 
         \SI{17.156}{\electronvolt} \\
         \SI{1.116}{\angstrom^3} & \SI{16.403}{\electronvolt} &
         \SI{16.401}{\electronvolt} & 
         \SI{16.387}{\electronvolt} \\
         \SI{1.259}{\angstrom^3} & 
         \SI{14.394}{\electronvolt} & 
         \SI{14.402}{\electronvolt} & 
         \SI{14.371}{\electronvolt} 
    \end{tabular}
    \caption{Fermi level computed with different smearing schemes and volumes 
    for
    the atomic Cs-IV phase. The electronic temperature is \SI{0.03}{\rydberg} and a $48 \times 48 \times 48$ grid has been used as a $\textbf{k}$-mesh. The error on the Fermi-level that 
    comes from the choice of the smearing scheme
    is lower than \SI{0.03}{\electronvolt}.
    }
    \label{tab:smearing}
\end{table}

\begin{figure}[b!]
    \centering
    \includegraphics[width=0.49\textwidth]{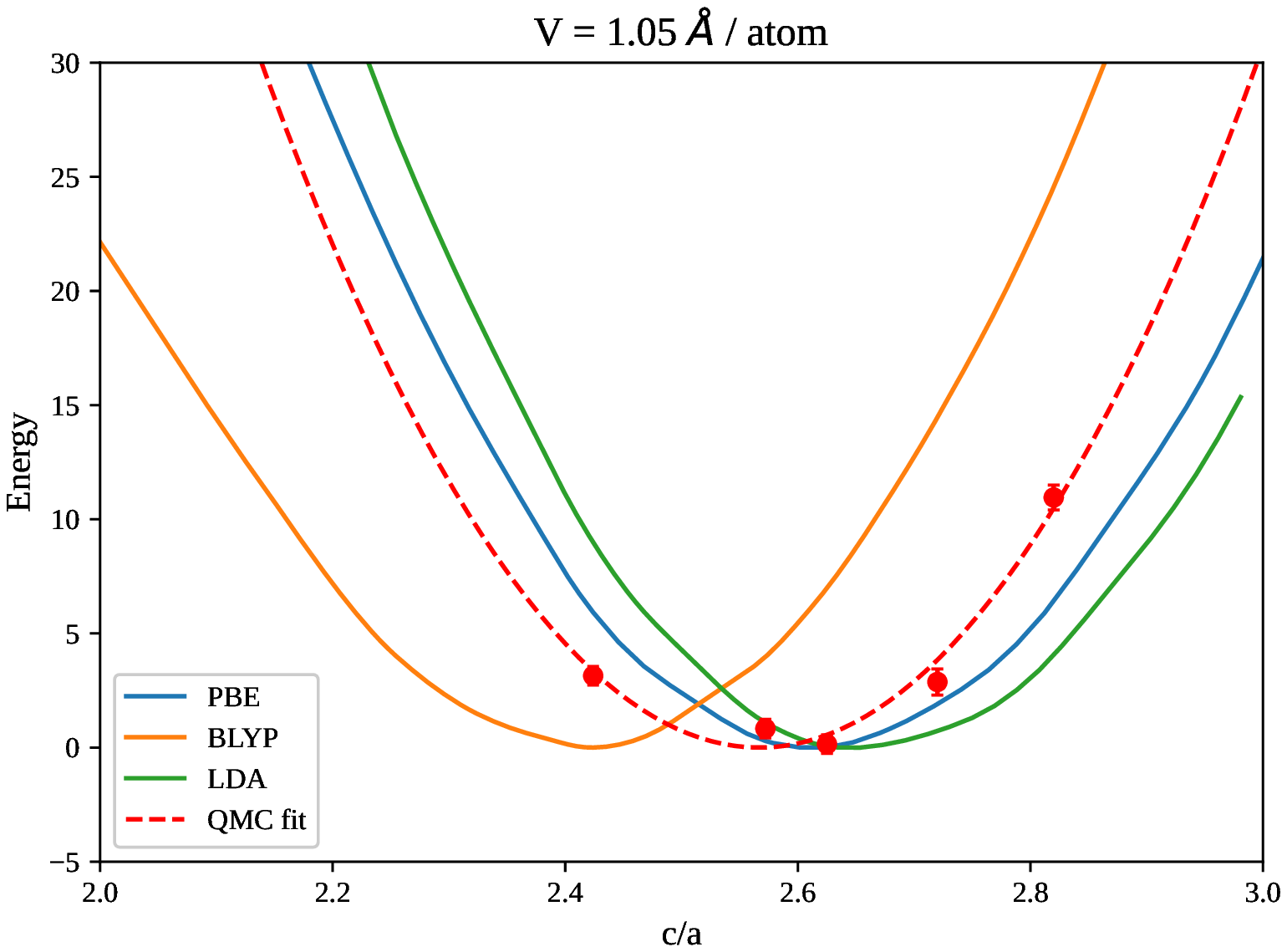}
    \includegraphics[width=0.49\textwidth]{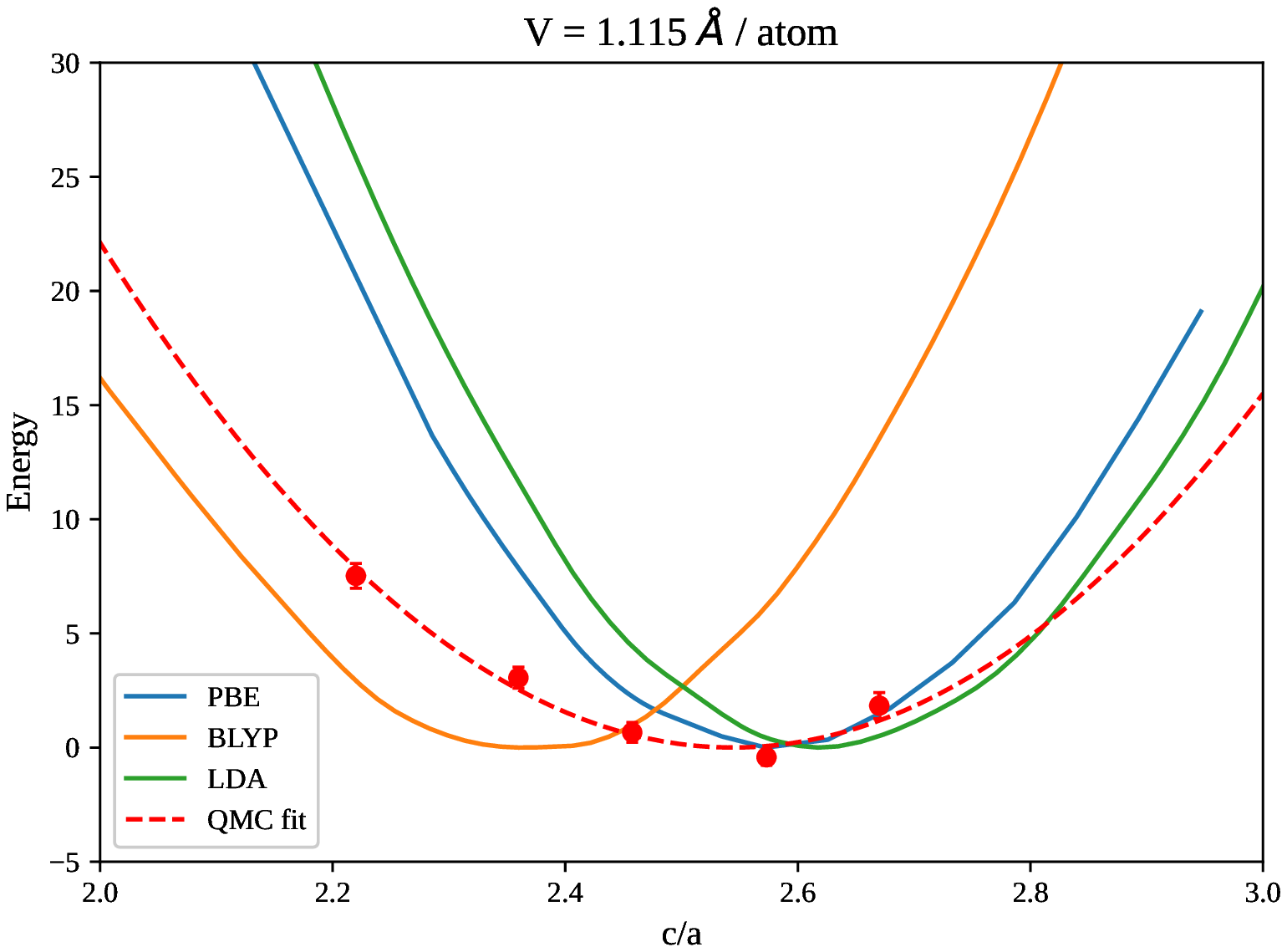}
    \includegraphics[width=0.49\textwidth]{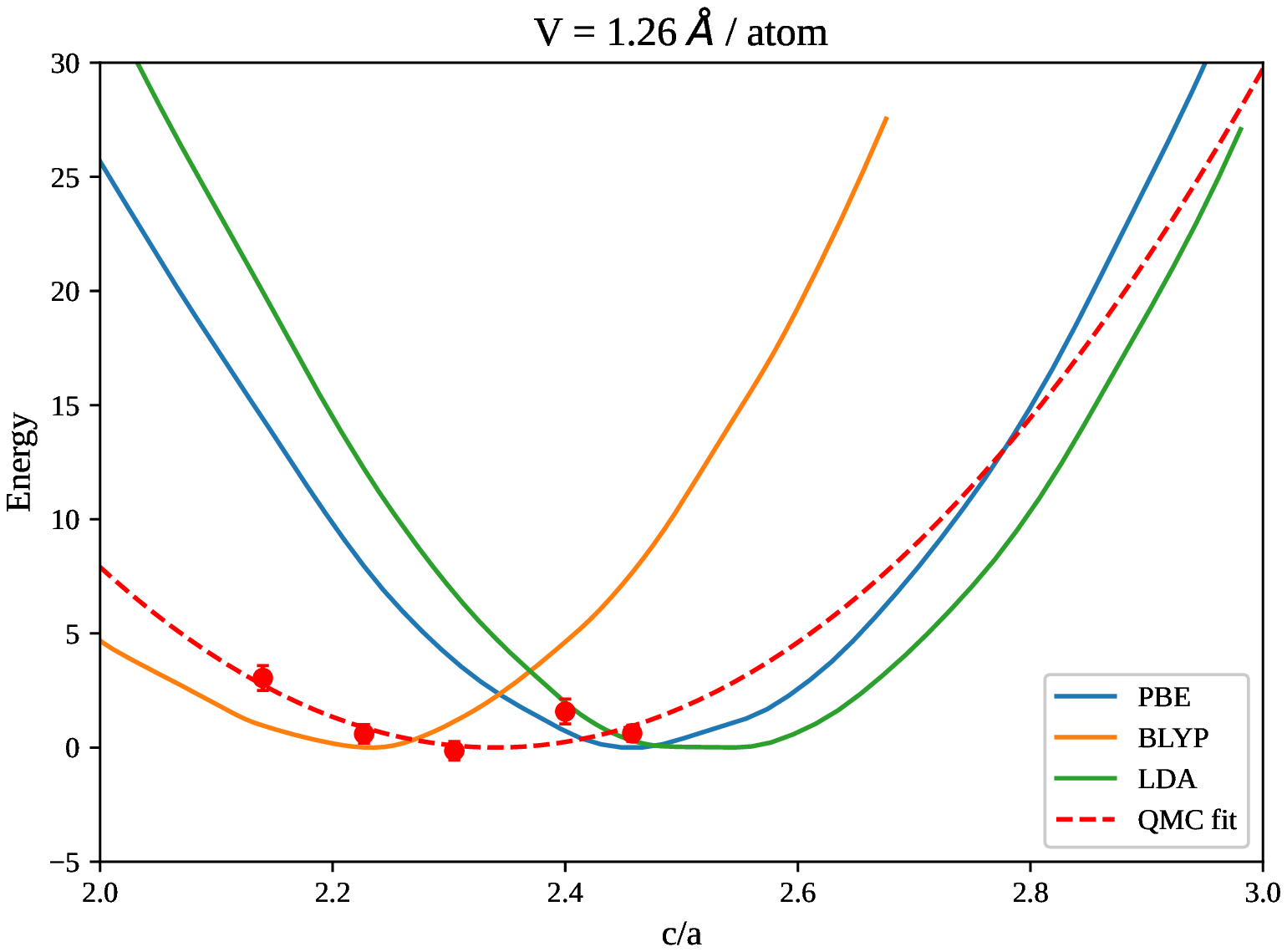}
    \caption{Energy profile as a function of the c/a parameter 
    for different volumes and electronic theories. Three of the c/a points taken in the DMC calculations are the static equilibrium value in BLYP, the SSCHA equilibrium value for deuterium and the one for protium.}
    \label{fig:energy:ca}
\end{figure}

The knowledge of
the actual volume dependence of the c/a equilibrium value is of paramount importance, because this 
could potentially have a strong impact on the electronic properties of the 
atomic
metallic 
phase of hydrogen. Indeed, 
the sudden increase of the DOS 
yielded
by the Lifshitz transition could affect the superconducting critical temperature. This is a fascinating scenario whose chance of occurrence needs to be addressed by a more accurate method such as QMC.
With the aim of determining the exact c/a equilibrium value,
we studied the c/a energy curve of BLYP, PBE, and LDA functionals, and compared them with reference results obtained by DMC calculations (\figurename~\ref{fig:energy:ca}).

From the comparison with the DMC results, 
one can 
notice 
that
the BLYP robustness
in providing accurate equilibrium geometry deteriorates at small volumes, corresponding to pressures above 560 GPa.
The PBE curves show instead the opposite behavior, as they become more and more accurate as the volume shrinks.
For the smallest volumes, the PBE results are the most accurate. Therefore, by looking at the right panel of Fig.~\ref{fig:lif}, which reports the PBE 
equilibrium c/a values compared with the critical c/a values for the Lifshitz transition, 
we can safely disregard the occurrence of 
this
transition in the pressure range where the atomic metallic phase becomes favorable. Indeed, nuclear quantum effects move the atomic phase further away from the transition.


\paragraph*{Optical properties}

We report here 
additional
information 
regarding
the optical properties of phase III, phase VI and the atomic one.

\begin{figure}[hbtp]
    \centering
    \includegraphics[width=\textwidth]{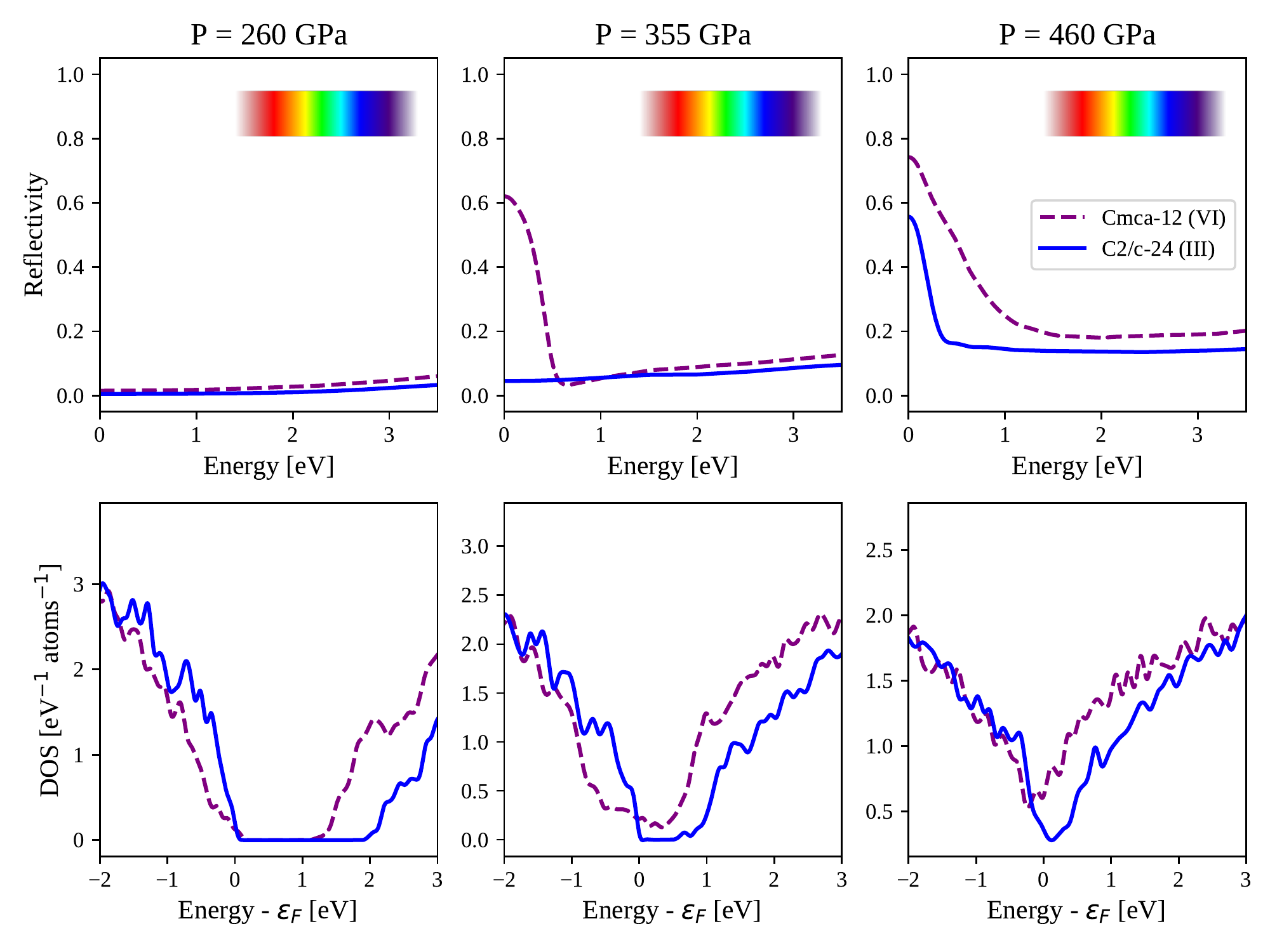}
    \caption{Reflectivity (upper row) and DOS (lower row) of phase III (C2/c-24) and phase VI (Cmca-12) at different pressures.}
    \label{fig:reflect}
\end{figure}

To complete the analysis presented in the main text about the
comparison between phase III and VI, 
in \figurename~\ref{fig:reflect} we show the electronic density of states (DOS) and the reflectivity of both phases.
As mentioned in the main text, phases III and VI are almost indistinguishable from reflectivity measurements in the visible range, but they present differences in the IR frequency range, where reflectivity is enhanced in phase VI. Also, the DOS at the Fermi level of phase VI is higher, resulting in a better DC conductivity of this phase, as reported in \figurename~\ref{fig:optic} of the main text.

To compute the optical properties of phase III and VI, we used supercells containing 324 atoms, where phonons are accounted for as static disorder (adiabatic approximation). We evaluated the refractive index and the transmitted light through a \SI{1.5}{\micro\meter} thick sample. This is the typical thickness of experimental samples at the target pressures. To avoid the systematic underestimation of the empty bands energy in the DFT calculation, we employed the modified Becke-Johnson meta-GGA exchange-correlation functional\cite{Tran_2009}, which is known to perform as well as more established (and more computationally expensive) methods like the HSE06 hybrid functional or the GW calculations\cite{Borlido_2019}.  All the calculations details, the equations employed and software used are the same as those discussed in the Methods Section of Ref.~\cite{MonacelliNatPhys2020}.

\begin{figure}[b!]
    \centering
    \includegraphics[width=0.8\textwidth]{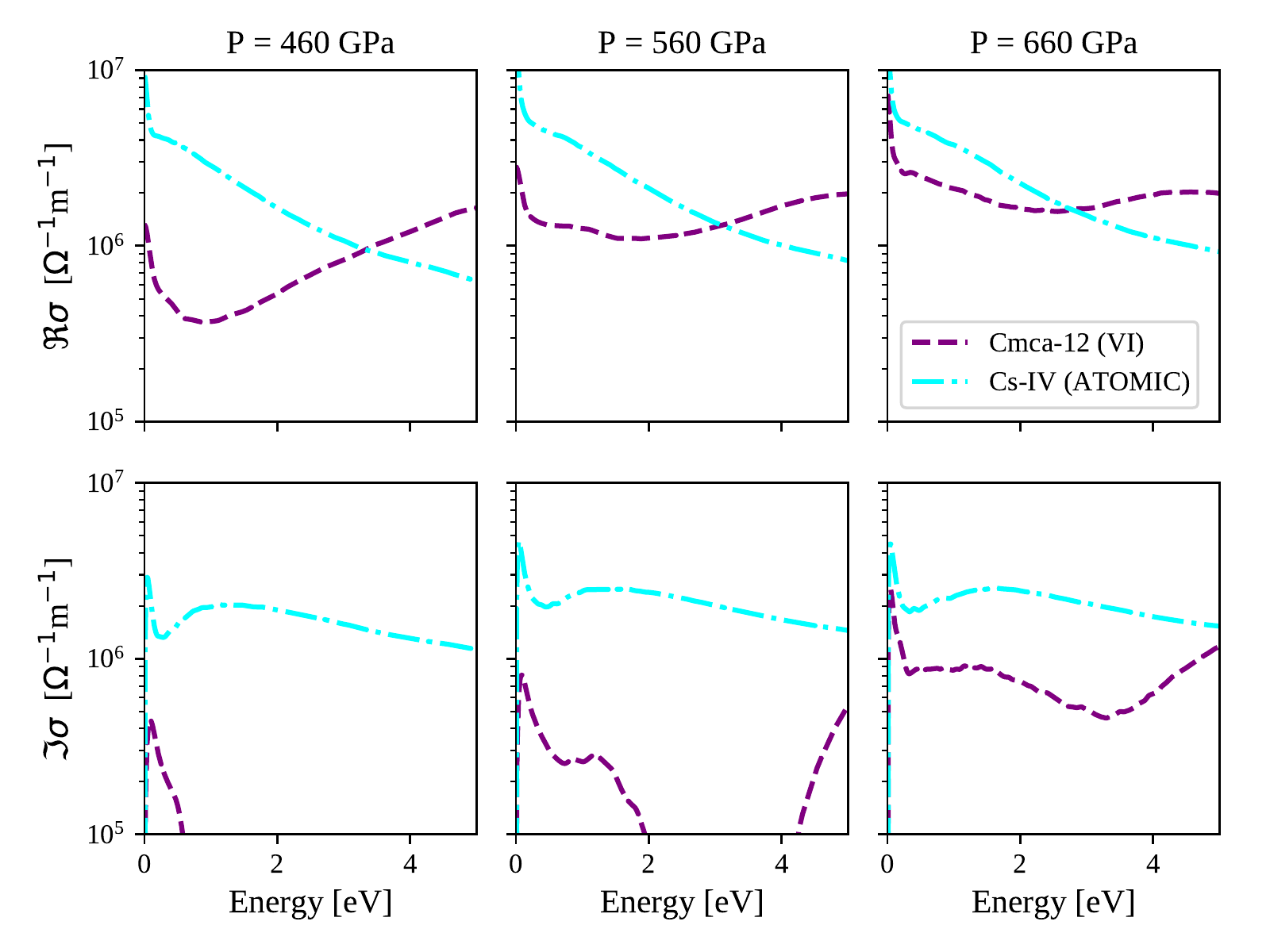}
    \caption{Real and imaginary part of the conductivity of Cmca-12 and Cs-IV phases.}
    \label{fig:conductivity}
\end{figure}

As far as the comparison between phase IV and the atomic phase is concerned,
we complement the data reported in the main text by including 
the optical conductivity (real and imaginary part) computed for the Cmca-12 and Cs-IV between 460 GPa and 660 GPa in \figurename~\ref{fig:conductivity}.

As 
mentioned in the paper,
our results on the optical properties are 
at variance
with the ones of Ref.~\cite{Gorelov2020}. This could be explained by the different approach used.
Indeed, in Ref.~\cite{Gorelov2020} the authors computed the optical properties on a 
smaller cell of 96 atoms and measured the optical gap by accounting 
for electron-hole pair excitations at the same point in the Brillouin zone of the 12-atom cell unfolded bands. In this way, 
scattering processes involving  phonon momenta $\textbf{q} \ne \Gamma$ are 
not included, 
where the excited electron-hole pair has a non-zero total momentum. In ordinary materials like silicon, these effects contribute as a small perturbation. However, this is not the case 
of
hydrogen, where the electron-phonon interaction shifts the bands by \SI{2}{\electronvolt}.

\begin{figure}[b!]
    \centering
    \includegraphics[width=0.85\textwidth]{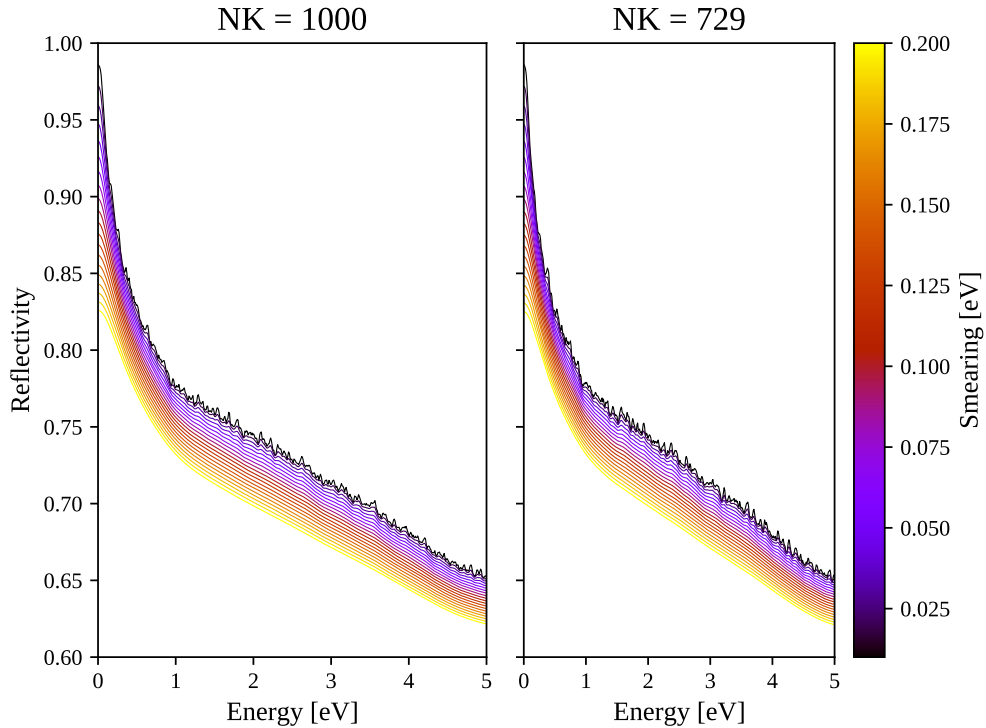}
    \caption{Convergence of the reflectivity as a function of smearing and number of $\textbf{k}$-points in the Cs-IV phase at approximately \SI{660}{\giga\pascal}.}
    \label{fig:smearing:r}
\end{figure}

\begin{figure}
    \centering
    \includegraphics[width=0.85\textwidth]{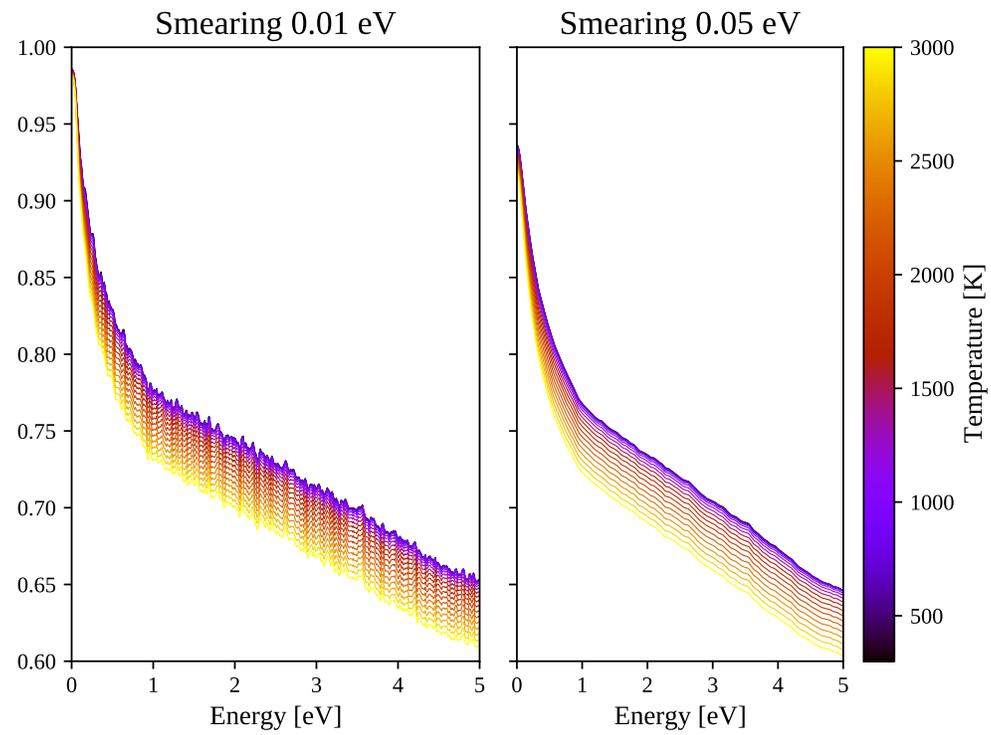}
    \caption{Convergence of the reflectivity as a function of the electronic temperature and 
    smearing
    in the Cs-IV phase at approximately \SI{660}{\giga\pascal}.}
    \label{fig:temperature:r}
\end{figure}

We carried out a thorough study of the convergence of the Cs-IV reflectivity with respect to the number of $\mathbf{k}$-points, the smearing (\figurename~\ref{fig:smearing:r}), and the electronic temperature (\figurename~\ref{fig:temperature:r}). The reflectivity shown in the main text has been obtained by employing a 729 $\mathbf{k}$-mesh with a smearing of \SI{0.05}{\electronvolt}.
The reflectivity depends 
weakly
on the electronic temperature when its value is lower than the smearing. Thus, 
at \SI{300}{\kelvin}, the temperature used in our analysis, the reflectivity is fully converged in its temperature dependence.


\newpage

\begin{figure}[b!]
    \centering
    \includegraphics[width=0.85\textwidth]{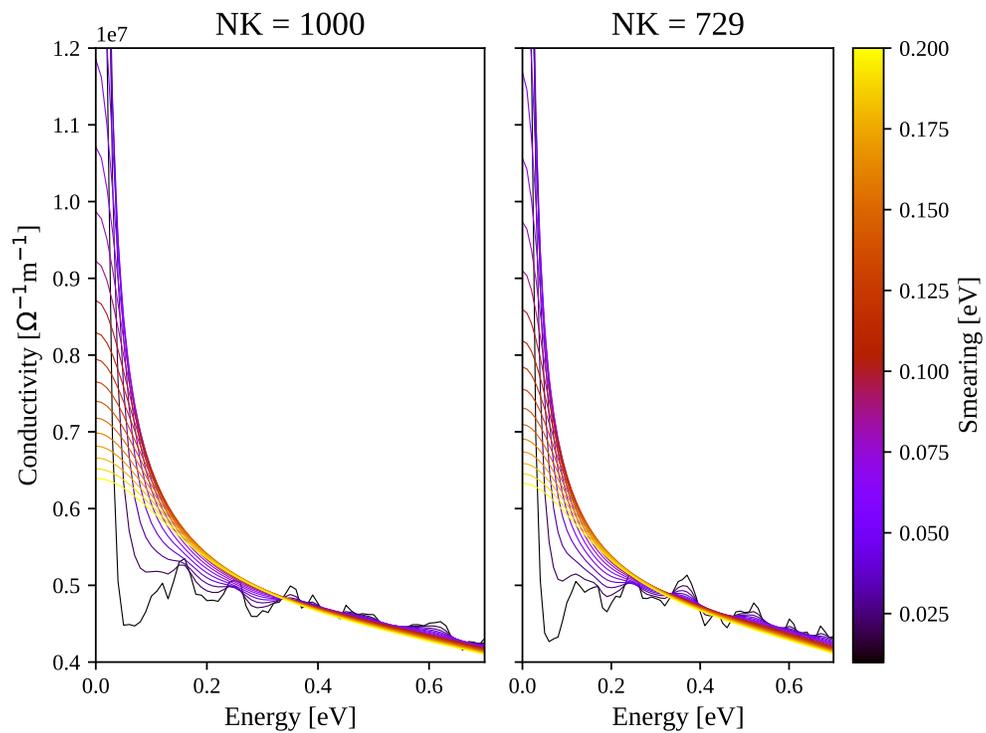}
    \caption{Convergence of the real conductivity as a function of smearing and number of $\textbf{k}$-points in the Cs-IV phase at approximately \SI{660}{\giga\pascal}.}
    \label{fig:optic:smearing}
\end{figure}

The most important dependence introduced by the finite smearing is the drop of reflectivity at low frequency. This is due to the strong but trivial dependence of the Drude peak. In \figurename~\ref{fig:optic:smearing} we show this effect, by plotting the real part of the conductivity as a function of smearing.

\newpage

\paragraph{Details on the DFT calculations}

\begin{figure}
    \centering
    \includegraphics[width=0.6\textwidth]{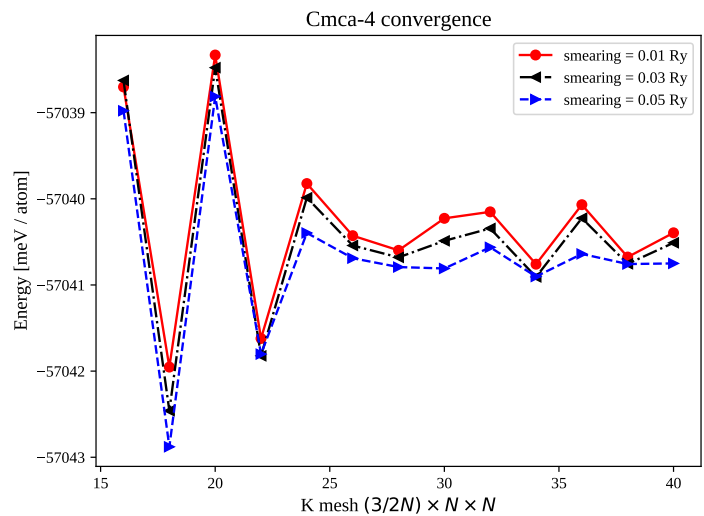}
    \caption{Molecular metallic Cmca-4 phase. DFT energy versus smearing and number of \textbf{k}-points in the primitive unit cell.}
    \label{fig:cmca4:kconv}
\end{figure}

\begin{figure}
    \centering
    \includegraphics[width=0.6\textwidth]{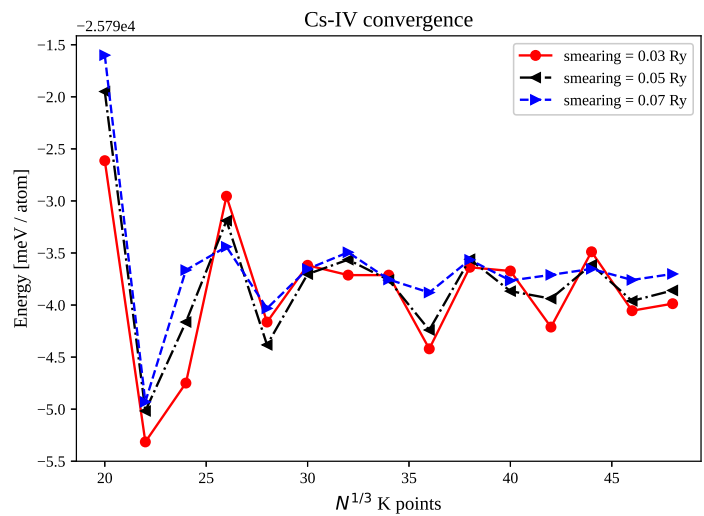}
    \caption{Atomic metallic Cs-IV phase. DFT energy versus smearing and number of \textbf{k}-points in the primitive unit cell.}
    \label{fig:csiv:kconv}
\end{figure}

For the DFT calculations we employed the Quantum Espresso\cite{Giannozzi2009,Giannozzi2017} software suite,
using a plane-wave basis set with a cutoff on the kinetic energy of \SI{1088}{\electronvolt} (\SI{4353}{\electronvolt} for the electronic density). We employed a norm-conserving pseudo-potential from the Pseudo Dojo library\cite{pseudodojo}.
To sample nuclear fluctuations within the SSCHA, the supercell contains 96 atoms for all the molecular structures (54 atoms for the atomic hydrogen with
finite-size convergence checked against a 128 atoms supercell). 
The electronic $k$ mesh is reported for each structure in \tablename~\ref{tab:my_label}. In all cases, a Marzari-Vanderbilt smearing of \SI{0.41}{\electronvolt} has been employed. Convergence of the energy with smearing and \textbf{k}-points is reported in \figurename~\ref{fig:cmca4:kconv} and \ref{fig:csiv:kconv} for the Cmca-4 and Cs-IV structures, respectively. The Cmca-4 and  Cs-IV phases are the ones with the most prominent metallic character, requiring the largest \textbf{k}-point sampling to converge. 

\begin{table}[h!]
    \centering
    \begin{tabular}{c|cc}
         & $\textbf{k}$-mesh \\
         \hline
         C2/c-24 & $12\times 12 \times 6$ \\
         P62/c-24 &  $12\times 12 \times 6$ \\
         Cmca-12 & $12\times 12\times 12$ \\
         Cs-IV & $48\times 48\times 48$ \\
         Cmca-4 & $36 \times 24 \times 24$ \\
    \end{tabular}
    \caption{\textbf{k}-mesh employed for the DFT simulation of each phase. The Cmca-4 mesh is performed in the conventional unit cell containing 8 atoms.}
    \label{tab:my_label}
\end{table}

\newpage

\paragraph{Details on the DMC calculations}

QMC calculations have been performed using the TurboRVB package\cite{nakano2020turborvb}. We carried out extensive DMC simulations in the lattice regularized DMC (LRDMC) flavor\cite{casula2005diffusion}, to project the initial many-body wave function towards the ground state of the system within the fixed-node approximation (FNA)\cite{anderson1976quantum}, and compute its energy. 

As starting many-body state, we employed a Jastrow-Slater variational wave function
$\Psi^\textbf{k}(\textbf{R})=\exp\{-U(\textbf{R})\} \det\{\phi^\textbf{k}_j(\textbf{r}^\uparrow_i)\}\det\{\phi^\textbf{k}_j(\textbf{r}^\downarrow_i)\}$ for $i,j \in \{1,\ldots,N/2\}$, where $N$ is the number of electrons in the unpolarized supercell, $\textbf{k}$ is the twist belonging to a Monkhorst-Pack (MP) grid of the supercell Brillouin zone, and $\textbf{R}=\{\textbf{r}^\uparrow_1,\ldots,\textbf{r}^\uparrow_{N/2},\textbf{r}^\downarrow_1,\ldots,\textbf{r}^\downarrow_{N/2}\}$ is the $N$-electron coordinate. 

$U$ is the Jastrow function, which is split into electron-nucleus, electron-electron, and electron-electron-nucleus parts: $U=U_{en}+U_{ee}+U_{een}$.
The electron-nucleus function has an exponential decay and it reads as $U_{en}=\sum_{iI} J_{1b}(r_{iI}) + U_{en}^\textrm{no-cusp}$, where the index $i$($I$) runs over electrons (nucleus), $r_{iI}$ is the electron-nucleus distance, and $J_{1b}(r)=\alpha (1-\exp\{-r/\alpha\})$, with $\alpha$ a variational parameter. $J_{1b}$ cures the nuclear cusp conditions, and allows the use of the bare Coulomb potential in our QMC framework. The electron-electron function has a Pad\'e form and it reads as $U_{ee}=-\sum_{i \ne j} J_{2b}(r_{ij})$, where the indices $i$ and $j$ run over electrons, $r_{ij}$ is the electron-electron distance, and $J_{2b}(r)=0.5 r/(1+ \beta r)$, with $\beta$ a variational parameter. This two-body Jastrow term fulfills the cusp conditions for antiparallel electrons. The last term in the Jastrow factor is the electron-electron-nucleus function: $U_{een}=\sum_{(i \ne j) I} \sum_{\gamma\delta} M_{\gamma \delta I}\chi_{\gamma I}(r_{iI}) \chi_{\delta I}(r_{jI})$, with $M_{\gamma \delta I}$ a matrix of variational parameters, and $\chi_{\gamma I}(r)$ a $(2s,2p,1d)$ Gaussian basis set, with orbital index $\gamma$, centered on the nucleus $I$. 
Analogously, the electron-nucleus cusp-free contribution to the Jastrow function, $U_{en}^\textrm{no-cusp}$, is developed on the same Gaussian basis set, such that  $U_{en}^\textrm{no-cusp}=\sum_{i I} \sum_{\gamma} V_{\gamma I}\chi_{\gamma I}(r_{iI})$, where $V_{\gamma I}$ is a vector of parameters.
The $J_{1b}$ and $J_{2b}$ Jastrow functions have been periodized using a $\textbf{r} \rightarrow \textbf{r}^\prime$ mapping that makes the distances diverge at the border of the unit cell, as explained in Ref~\cite{nakano2020turborvb}. For the inhomogeneous $U_{een}$ part, the Gaussian basis set $\chi$ has been made periodic by summing over replicas translated by lattice vectors.

The one-body orbitals $\phi$ are expanded on a primitive $(4s,2p,1d)$ Gaussian basis set, which we contracted into 6 hybrid orbitals, by using the geminal embedding orbitals (GEO) contraction scheme\cite{sorella2015geminal} at the $\Gamma$ point. $\phi$s' are made periodic by using the same scheme as for the $\chi$s'. We verified that this basis set yields a FN-LRDMC bias in the energy differences smaller than the target error of 1 meV per atom. 
For each $\textbf{k}$ belonging to the MP grid of a given supercell, we performed independent DFT calculations in the local density approximation (LDA) to generate $\{\phi^\textbf{k}_j\}_{j=1,\ldots,N/2}$ for all occupied states. Note that these LDA calculations are done for an \emph{ab initio} Hamiltonian with bare Coulomb potential for the electron-ion interactions. This is thanks to the one-body Jastrow factor included in the DFT wave function,
with $\alpha \approx 2.5$. In presence of a Coulomb divergence, fulfilling the ion cusp conditions accelerates enormously the basis set convergence already at the DFT level.

Before running LRDMC calculations, we optimized the $\alpha$, $\beta$ and $M_{\gamma \delta I}$ parameters, by minimizing the variational energy of the wave function $\Psi$ within the QMC linear optimization method\cite{umrigar2007alleviation}, by keeping the orbitals $\phi^\textbf{k}_i$ fixed. All $\textbf{k}$ twists belonging to the same system share the same set of optimal variational parameters for the Jastrow factor. The LRDMC projection is carried out at the lattice space $a=0.25 a_0$, which yields converged energy differences. The projection algorithm has been implemented with a fixed population of 256 walkers per twist for the largest system sizes. The population bias, falling within the error bars, has been corrected by the ``correcting factors'' scheme\cite{buonaura1998numerical}.

\begin{figure}[th!]
    \centering
    \includegraphics[width=0.49\textwidth]{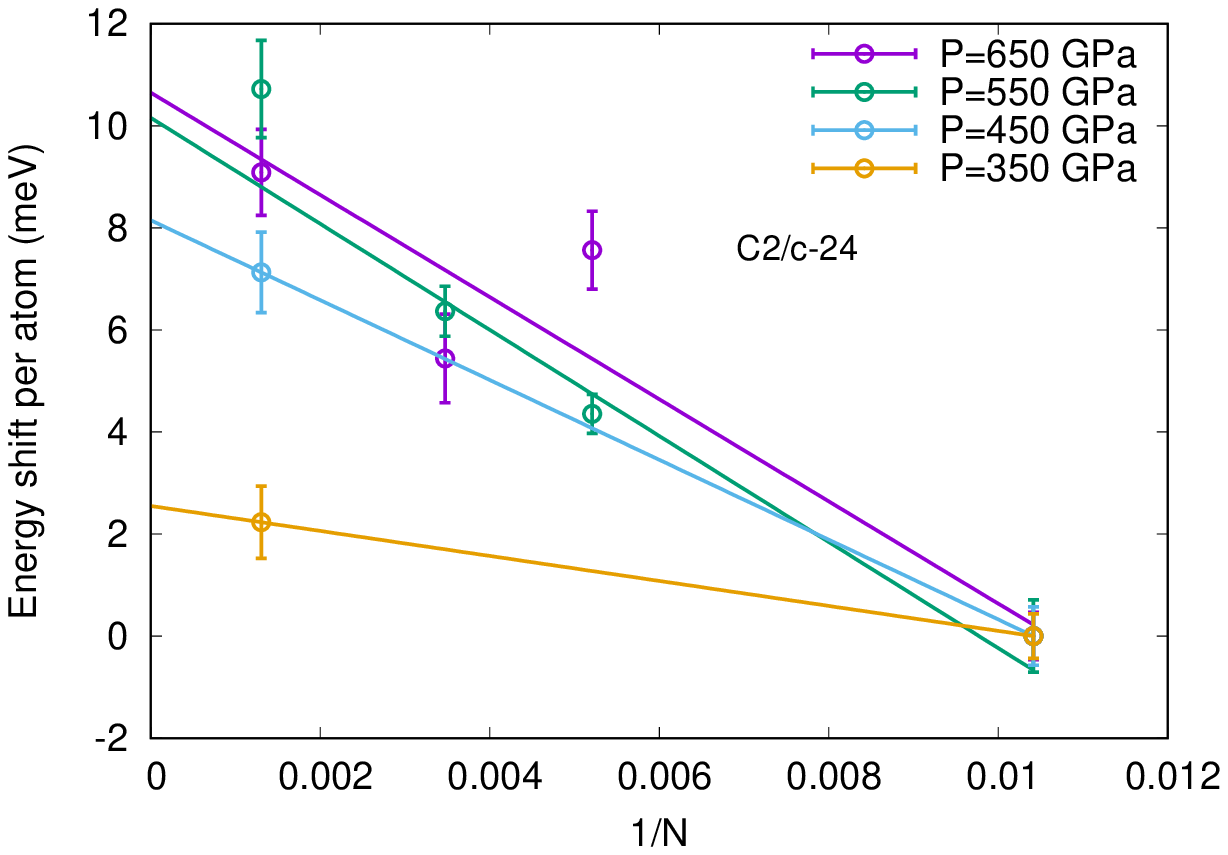}
    \includegraphics[width=0.49\textwidth]{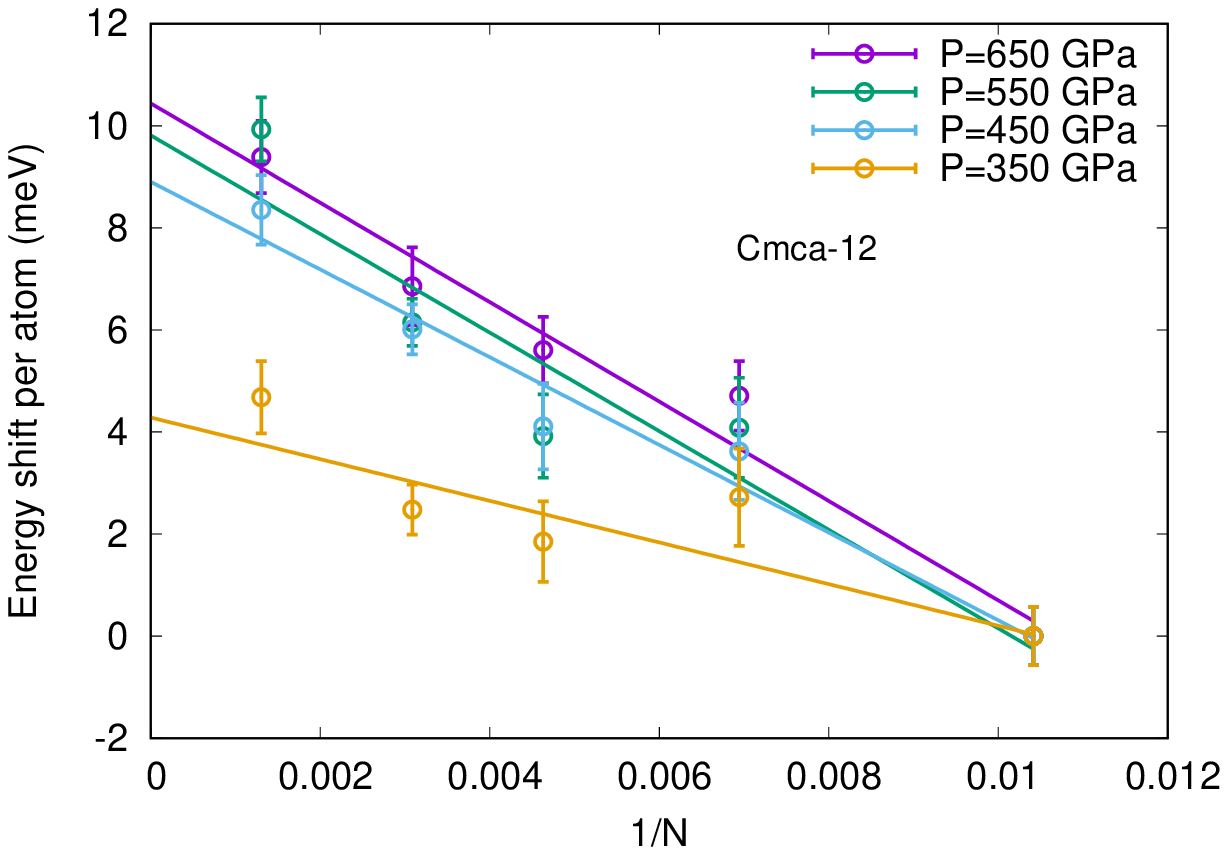}
    \includegraphics[width=0.49\textwidth]{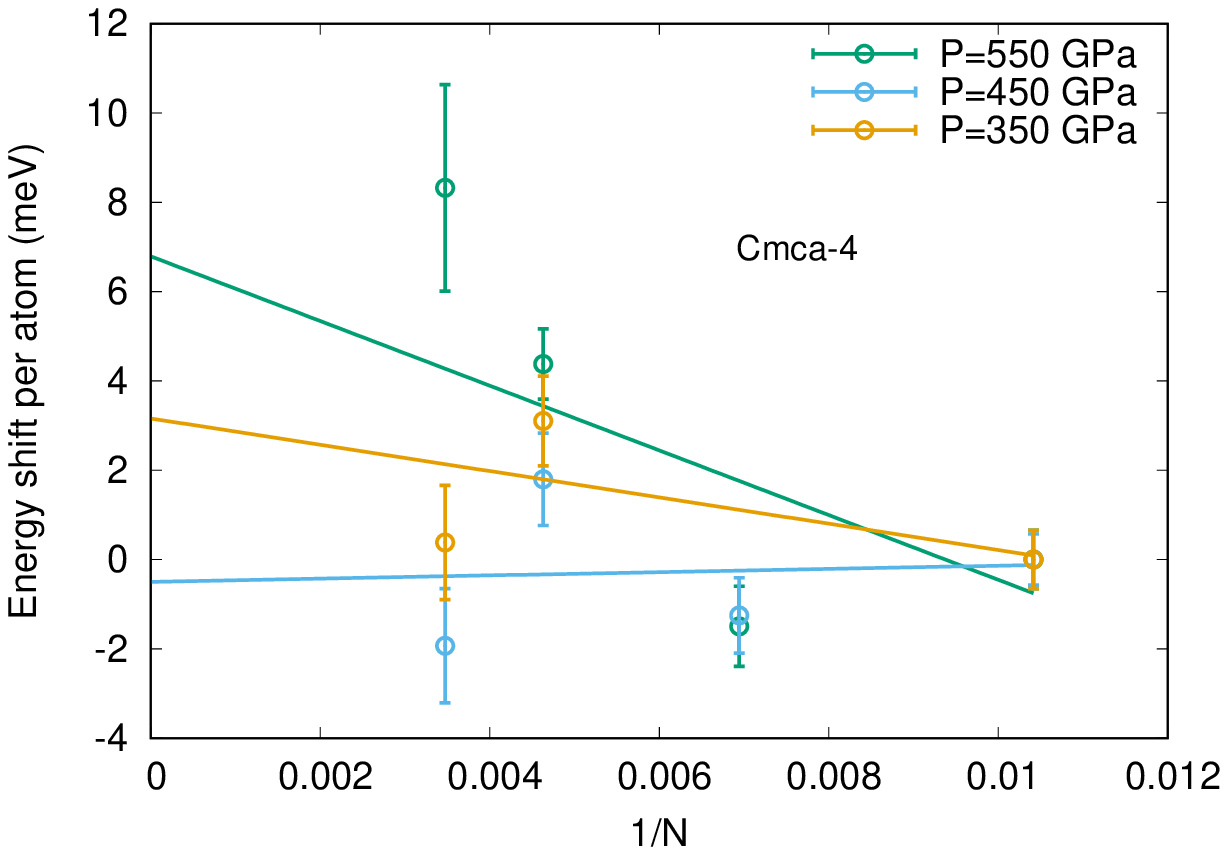}
    \includegraphics[width=0.49\textwidth]{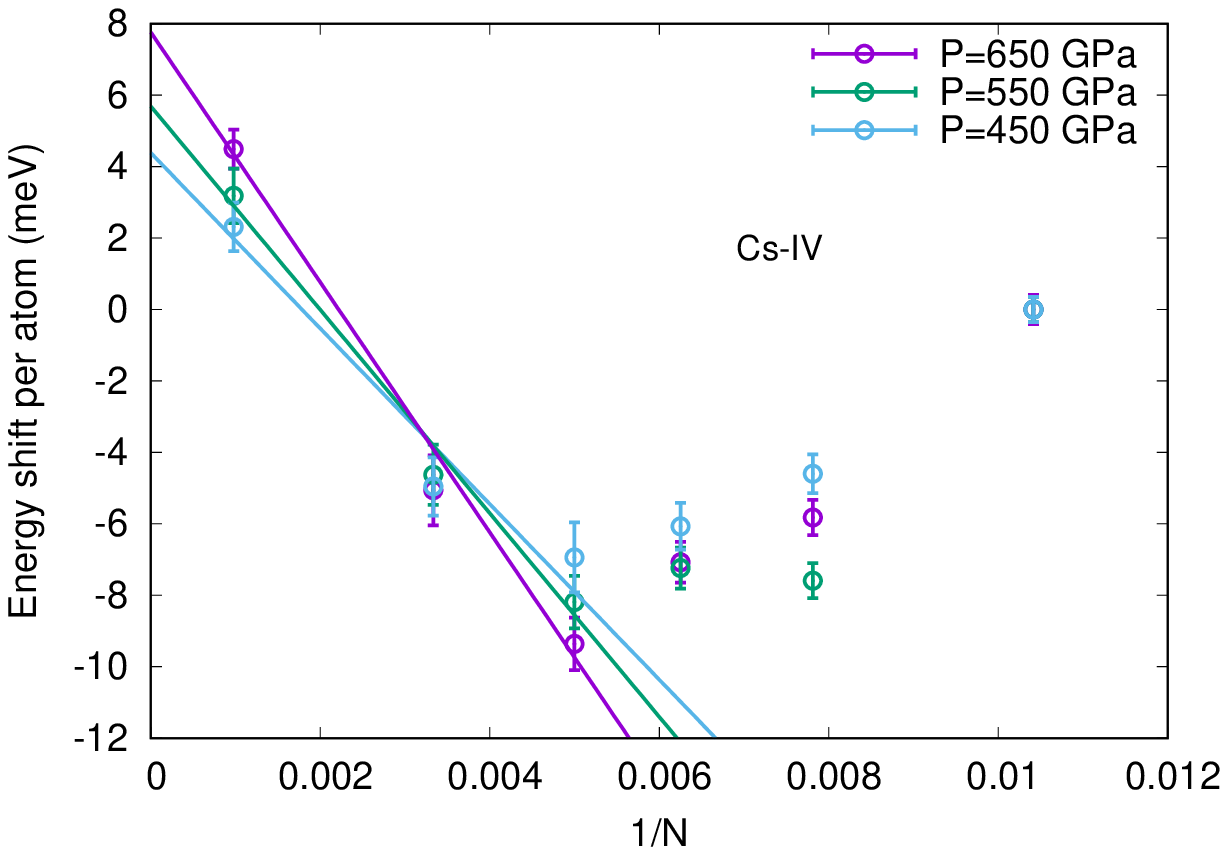}
    \caption{QMC finite-size scaling and extrapolation to the thermodynamic limit. KZK-corrected LRDMC energies for 4 crystalline symmetries (C2/c-24, Cmca-12, Cmca-4, and Cs-IV) plotted as a function of $1/N$, where $N$ is the number of atoms, with respect to their value at $N=96$, taken as reference. The energies are twisted-averaged in the canonical ensemble over a $\textbf{k}$-grid that has been rescaled according to the size of the supercell, as explained in the text. Note that despite the KZK correction and the canonical $\textbf{k}$-average, there is a residual size dependence beyond $N=96$, larger than the target accuracy of 1 meV per atom, that needs to be extrapolated. As expected, this residual dependence is stronger in the atomic metallic phase and in the molecular phases under high pressure, where the metallic character is enhanced.}
    \label{fig:qmc:finite-size-extrapolation}
\end{figure}

For each lattice symmetry and volume $V$, we performed a size-scaling analysis to extrapolate the energies to the thermodynamic limit (see Fig.~\ref{fig:qmc:finite-size-extrapolation}). Let $N_x\times N_y\times N_z$ be the electronic $\textbf{k}$-mesh yielding converged DFT results. In QMC, we used the same $\textbf{k}$-meshes reported in Tab.~\ref{tab:my_label}, except for the Cs-IV and Cmca-4 symmetries, where we used a slightly smaller $24\times24\times24$ and $18\times12\times12$ mesh, respectively. To further reduce finite-size errors, the $\textbf{k}$-mesh of the metallic Cs-IV symmetry has been centered at $(\pi, \pi, \pi)$, while the other $\textbf{k}$-grids are centred at $\Gamma$. We then took supercells with volume $V_s= L_x L_y L_z V$, where $L_i$ are the number of unit-cell replica in the $i$-th direction. Accordingly, the twists have been taken as belonging to the $M_x\times M_y\times M_z$ MP $\mathbf{k}$ mesh with $M_i= \textrm{int} [ N_i / L_i ]$, where $\textrm{int}$ is the integer function. The ground state energies have been extrapolated by using supercells as large as $N=768$ for the molecular phases, while for atomic I4/amd-2 symmetry we used supercells as large as $N=1024$. The final extrapolations have been performed by a linear fitting in $1/N$, computed with Kwee-Zhang-Krakauer (KZK)-corrected energies\cite{kwee2008finite}.

We also investigated the role of the FNA in DMC, by exploiting the capabilities of TurboRVB for optimizing the Slater orbitals $\phi^\textbf{k}_i$. Due to the increased cost of these simulations, we performed the nodal optimization by the variational Monte Carlo energy minimization at the special $\textbf{k}$-point only\cite{dagrada2016exact}. The corresponding DMC energies computed by projecting wave functions with relaxed nodes prove that the FN bias
does not affect the relative energies between molecular phases. However, the optimization of the FNA with respect to the LDA nodes in the QMC wave function shifts the atomic-phase energy upwards by about \SI{3}{\milli\electronvolt} per atom (see Tab.~\ref{tab:FN_label}), also shifting the transition pressure by \SI{20}{\giga\pascal} towards higher pressures (the correct value is accounted for in the phase diagram of \figurename~\ref{fig:PD}).

\begin{table}[h!]
    \centering
    \begin{tabular}{c | c | c | c}
         symmetry & 1.416 \AA$^3$ & 1.259 \AA$^3$ & 1.115 \AA$^3$ \\
         \hline
         C2/c-24 &  6.2 ($\pm$ 1.7) & - & - \\
         P62/c-24 & 6.0 ($\pm$ 1.7) & - & - \\
         Cmca-4 &  5.9 ($\pm$ 1.5) & - & - \\
         Cmca-12 & 7.2 ($\pm$ 1.1) &  6.5 ($\pm$ 1.1) & 6.1 ($\pm$ 1.0)\\
         Cs-IV &  -  & 3.4 ($\pm$ 0.8) & 3.2 ($\pm$ 0.8)  \\
    \end{tabular}
    \caption{Fixed node LRDMC energy gain (in meV/H) at different volumes with respect to the LDA nodes after full wave function optimization at the special $\textbf{k}$ point. The energy optimization has been performed at the variational Monte Carlo level for supercells up to $N=288$. It turns out that the LRDMC energy gain due to the nodal optimization has a very weak system size dependence. $N=96$ gives already converged results for the FN energy gain.}
    \label{tab:FN_label}
\end{table}

We ran all LRDMC calculations long enough to reach a stochastic error bar around the target accuracy of \SI{1}{\milli\electronvolt} per atom.

\newpage

\paragraph*{DMC correction of the DFT exchange-correlation energy}

\begin{figure}
    \centering
    \includegraphics[width=\columnwidth]{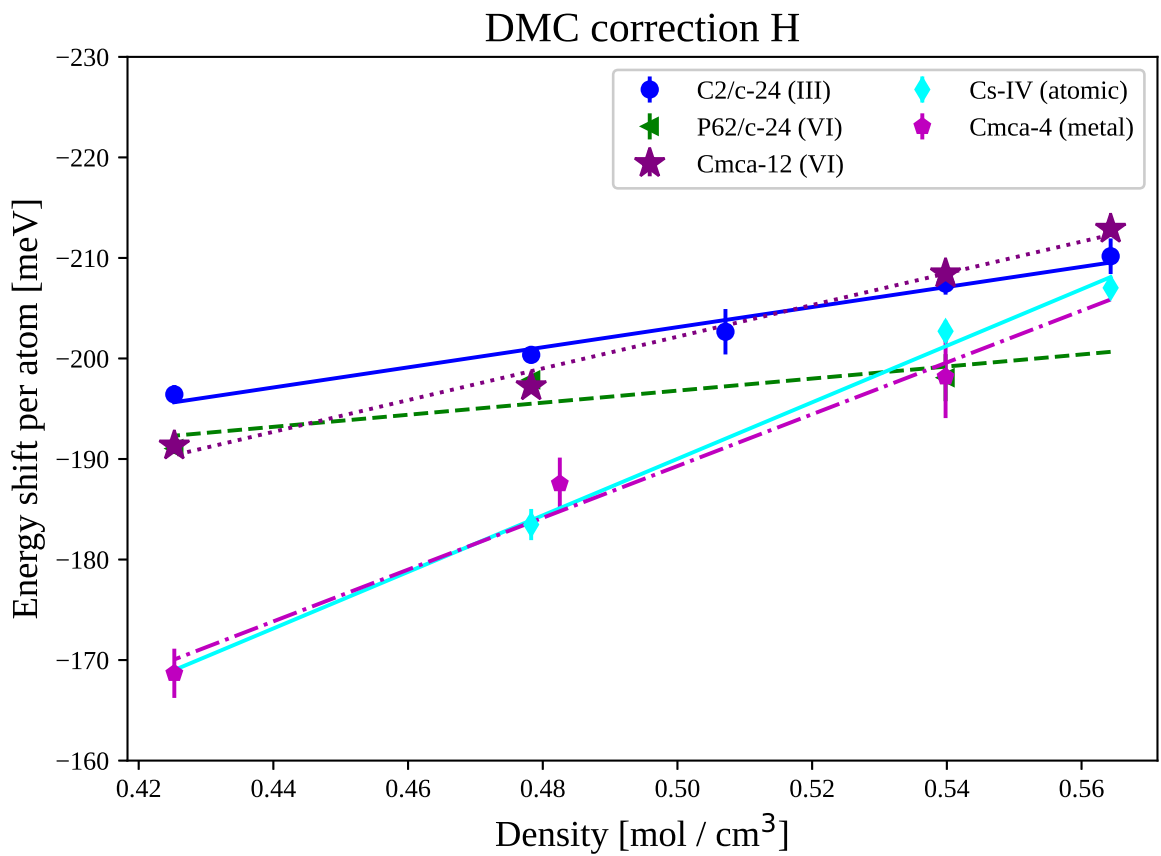}
    \caption{QMC energy corrections. Electronic energy differences between DFT-BLYP and 
    DMC calculations at hydrogen centroid positions obtained from SSCHA nuclear quantum fluctuations evaluated at the DFT-BLYP level. The DMC energies are computed within the FNA with DFT-LDA nodes. 
    The fit is a straight line for all phases.
    }   
    \label{fig:qmcshift}
\end{figure}

To correct for the DFT exchange-correlation error, we computed DMC energies at each centroids structure.
For each structure, we fitted the energy difference between DFT and 
DMC for the same structure as a function of the density, 
and added it on the top of the DFT energy-volume relationship, computed on a much denser volume grid thanks to the cheaper cost of DFT. With this procedure, we do not rely on any phenomenological definition of the equation of states (EOS), such as the Birch-Murnaghan or Vinet EOS, in order to get QMC-interpolated energy-versus-volume curves. Fitting the DMC \emph{corrections} with respect to an underlying \emph{ab initio} theory is easier than fitting directly the DMC total energies. Indeed, total energies show a much larger dependence on the volume than energy corrections. 
The plot of the QMC corrections is reported in \figurename~\ref{fig:qmcshift}. In this plot, the QMC corrections are obtained from DMC energies computed within the fixed-node approximation (FNA) and with DFT-LDA nodes. 

Fig.~\ref{fig:qmcshift} shows that the difference between phases smears out when the density increases, pointing toward a
better DFT description above densities corresponding to \SI{800}{\giga\pascal}, where the electronic behavior of the system is more similar to the jellium model. This regime, however, kicks in at pressures above the range of interest for this work. It is worth noting that the smallest absolute values of the DMC correction with respect to DFT-BLYP are found for the atomic and Cmca-4 phases. This is the reason why accounting for DMC corrections is fundamental to correctly reproduce the hydrogen atomization: all molecular phases, except for Cmca-4, are lowered in energy with a consequent shift of the atomization transition toward higher pressures.

We have mentioned that the QMC corrections reported in \figurename~\ref{fig:qmcshift} are based on DMC calculations within the FNA and with DFT-LDA nodes. As explained in the QMC calculations details, we assessed the quality of the DFT-LDA wave-function nodes that are kept fixed during the DMC projection in the FNA. This is the only bias present in the DMC energies, which would otherwise have been exact. We have been able to relax the nodes at the variational Monte Carlo level, and then use improved nodes in DMC. According to Tab.~\ref{tab:FN_label}, a systematic gain of 3meV/H is found after nodal optimization in DMC energies for the molecular phases with respect to the atomic one. It appears that the gain is the same for all molecular structures (within the error bars) and it is volume independent in the range of pressures explored in this work. This 3meV/H shift is added to the QMC corrections in Fig.~\ref{fig:qmcshift} to yield the final corrections, used to compute the QMC phase diagram of \figurename~\ref{fig:PD}. It adds up to further disfavor the atomic phase, whose stability is pushed up to higher pressures in the phase diagram.

\paragraph*{DMC+SSCHA systematic errors}

Our approach of combining DMC energy corrections and SSCHA anharmonic vibrational contributions in an additive way relies upon the hypothesis that DMC corrections depend 
mainly
on the phase and pressure (or volume), 
and very weakly
on the particular atomic displacement around the centroids within a given phase. To test this hypothesis and give an estimate of the systematic error introduced by this approximation, we repeated the same calculations by choosing a different reference structure to compute the DMC shifts. Therefore, we replaced the SSCHA centroids of hydrogen with those of deuterium to change reference structure, and we went again through all steps by using this time the DMC correction computed on the D centroids, for validation and error quantification. 

\begin{figure}
    \centering
    \includegraphics[width=\textwidth]{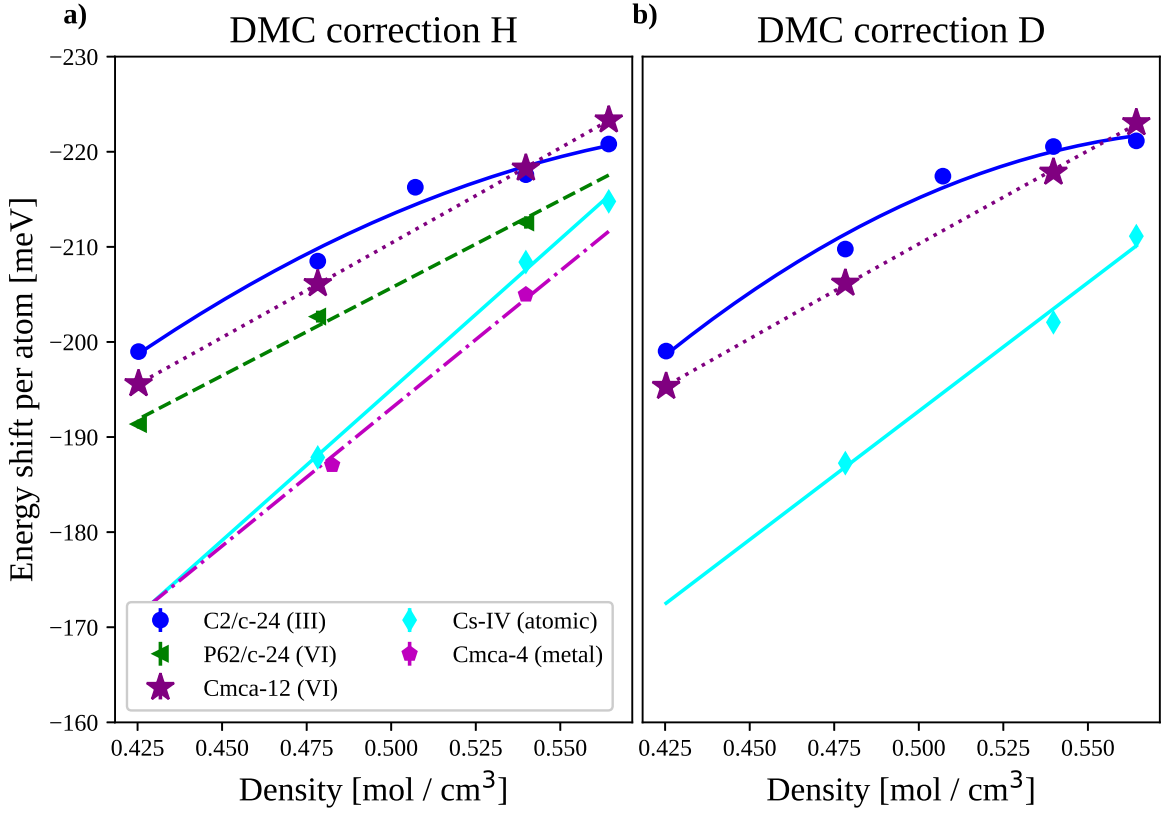}
    \caption{Electronic energy difference per atom between DMC and
    DFT calculations on the same BLYP structures. Here, we used the KZK-corrected LRDMC energies computed at $N=96$ for all phases reported in the plot, before full thermodynamic extrapolation. This allows for a direct comparison with deuterium, where we did not perform an explicit finite-size extrapolation as we did in hydrogen (see Fig.~\ref{fig:qmc:finite-size-extrapolation}).
    Panel a: the energy shift per atom is computed on the SCHA centroid positions with mass equal to hydrogen. Panel b: the energy shift is computed on structures with SCHA centroid positions for deuterium.}
    \label{fig:qmcshift:96}
\end{figure}

In \figurename~\ref{fig:qmcshift:96}, we compare the DMC corrections computed for the SSCHA centroids of $\text{H}$ and $\text{D}$, respectively, employing supercells of 96 nuclei in the DMC calculations. In this way, we can 
check whether the DMC shifts are independent of specific ionic distortion, by retaining a dependence only on the overall arrangement, i.e. on the crystalline symmetry. The plots in panels a) and b) show that the shift between C2/c-24 and Cmca-12 depends only slightly on the centroid position, while the atomic phase is more 
sensitive to the choice of the centroid. This discrepancy is mainly due to the difference in the c/a equilibrium values of Cs-IV
between DMC and DFT-BLYP (see Section on the Cs-IV phase).

\newpage

\begin{figure}
    \centering
    \includegraphics[width=.7\textwidth]{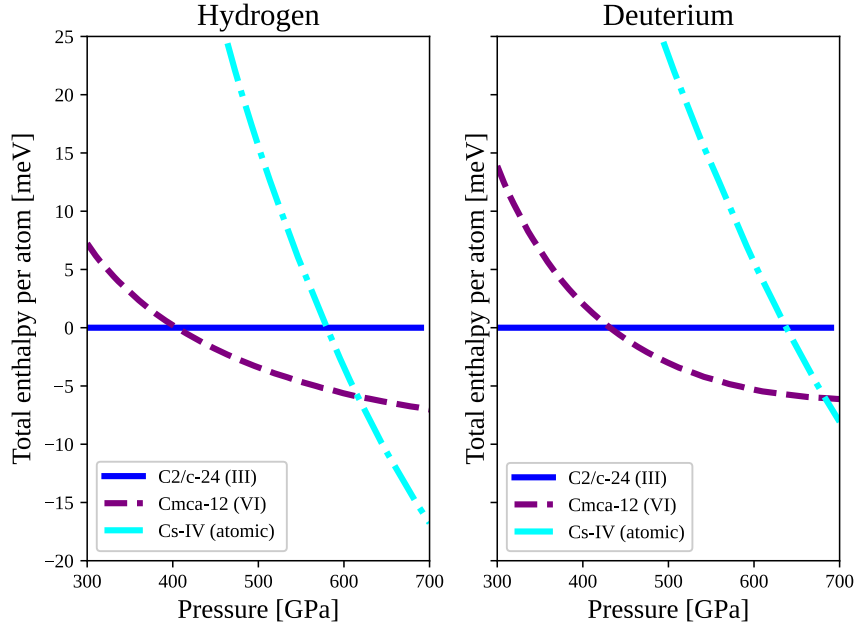}
    \caption{Full phase diagram, considering DMC corrections and nuclear quantum effects. DMC corrections are evaluated using the SSCHA centroids of deuterium.
    }
    \label{fig:pd:full:D}
\end{figure}

In \figurename~\ref{fig:pd:full:D}, we show the phase diagram obtained by considering the DMC correction on the $\text{D}$ centroids. Here, the DMC energies computed on the $\text{D}$ centroids are extrapolated to the thermodynamic limit by assuming the same $1/N$ dependence as found in protium. Therefore, the most accurate DMC correction of the BLYP exchange-correlation energy is obtained for the H-centroids geometries, where an explicit and computational time-consuming extrapolation to the thermodynamic limit has been performed explicitly. Nonetheless, the phase diagram in
\figurename~\ref{fig:pd:full:D}
is 
valuable for
estimating the systematic errors. This can be done by direct comparison with the phase diagram drawn in 
\figurename~\ref{fig:pd:full},
obtained instead by relying upon the H-centroids DMC correction.


Based on the phase diagrams plotted in Figs.~\ref{fig:pd:full} and \ref{fig:pd:full:D}, 
we report the phase-transition pressures for hydrogen and deuterium in Tabs.~\ref{tab:PD:H} and \ref{tab:PD:D}, respectively. The discrepancies between the results is used as an estimate of the systematic error in the final phase diagram, shown in \figurename~\ref{fig:PD}. 

\begin{table}
\centering
\begin{tabular}{c|c|c}
&III $\rightarrow$ VI & VI $\rightarrow$ ATOMIC\\
\hline
$\text{H}$-centroids DMC correction & \SI{422}{\giga\pascal} & \SI{575}{\giga\pascal} \\
$\text{D}$-centroids DMC correction & \SI{404}{\giga\pascal} & \SI{616}{\giga\pascal} 
\end{tabular}
\caption{The final phase-transition pressures for H obtained with DMC energy correction estimated based on H or D centroids. 
The results obtained from the H-centroids based DMC correction are taken as the 
most
accurate in the main text. The ones 
obtained from the D-centroids based DMC correction 
provide
an estimation of systematic error introduced by the approximation.
\label{tab:PD:H}}
\end{table}

\begin{table}
\centering
\begin{tabular}{c|c|c}
&III $\rightarrow$ VI & VI $\rightarrow$ ATOMIC\\
\hline
$\text{H}$-centroids DMC correction & \SI{452}{\giga\pascal} & \SI{646}{\giga\pascal} \\
$\text{D}$-centroids DMC correction & \SI{432}{\giga\pascal} & \SI{683}{\giga\pascal} 
\end{tabular}
\caption{Same transition pressures as in Fig.~\ref{tab:PD:H}, but for deuterium.
The results obtained from the H-centroids based DMC correction are taken as the 
most
accurate in the main text. The ones 
obtained from the D-centroids based DMC correction 
provide
an estimation of systematic error introduced by the approximation.\label{tab:PD:D}}
\end{table}

\clearpage

\end{document}